\documentclass[preprint,10pt,times,onecolumn]{elsarticle}
\usepackage[%showframe,%Uncomment any one of the following lines to test 
margin=1.5cm,
]{geometry}
\usepackage{amsmath}
\usepackage{amsfonts}
\usepackage{amssymb}
\usepackage{graphicx}
\usepackage{lmodern}
\usepackage{hyperref}
\usepackage{dcolumn}% Align table columns on decimal point
\usepackage{bm}% bold math
\usepackage[mathlines]{lineno}% Enable numbering of text and display math
\usepackage{float}
\usepackage[T1]{fontenc}
\usepackage{subfigure}
\usepackage{natbib}
\usepackage{breqn}
\allowdisplaybreaks
\journal{Physics Letters B}
\usepackage{numcompress}%\bibliographystyle{model6-num-names}

\DeclareUnicodeCharacter{2212}{\ensuremath{-}}
\begin{document}
\newcolumntype{P}[1]{>{\centering\arraybackslash}p{#1}}
\title{\textbf{ The Echo 12–23 Texture: A Novel Flavour Paradigm for Neutrinos}}
%\tnotetext[mytitlenote]{Fully documented %templates are available in the elsarticle %package on \href{http://www.ctan.org/tex-%archive/macros/latex/contrib/elsarticle}{CTAN}.}
\author[aff1]{Manash Dey\corref{cor1}\fnref{present1}}
\cortext[cor1]{Corresponding author}
\fntext[present1]{Present affiliation: Department of Physics, Maibang Degree College, India, 788831.}

\ead{manashdey@gauhati.ac.in; manashdey1272@gmail.com}

\author[aff1]{Subhankar Roy}
\ead{subhankar@gauhati.ac.in}

\address[aff1]{Department of Physics, Gauhati University, India, 781014.}

% E-mail addresses: for both authors
\begin{abstract}
We construct a flavour guided model that realises the distinctive \emph{Echo 12–23 Texture} in the neutrino mass matrix through a non-trivial interplay of symmetries. While the model accounts for charged lepton mass hierarchy, the texture offers interesting insights specifically into neutrino mass ordering and flavour dynamics. With clear imprints on low energy observables, the framework provides a minimal yet testable path toward understanding the origin of neutrino properties.
\end{abstract}
\maketitle
\flushbottom
%\tableofcontents
\section{Introduction}
\label{section 1}
\begin{figure}[h!]
    \centering
    \begin{minipage}[b]{0.25\textwidth}
        \centering
        \includegraphics[width=\textwidth]{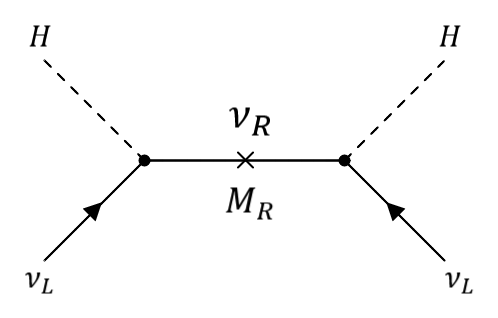}
        (a) Type-I
    \end{minipage}
    \hspace*{1cm}
    \begin{minipage}[b]{0.15\textwidth}
        \centering
        \includegraphics[width=\textwidth]{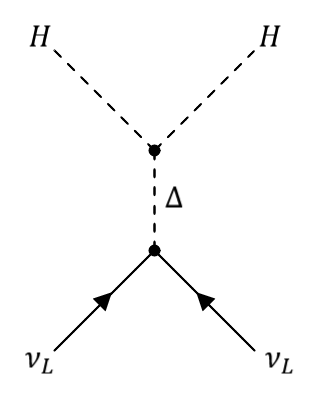}
        (b) Type-II
    \end{minipage}
    \caption{Feynman diagrams for Type-I and Type-II seesaw mechanisms.}
\label{fig:feynman}
\end{figure}

The Standard Model (SM)\,\cite{Glashow:1961tr, Weinberg:1967tq, Salam:1968rm, Gross:1973id, Politzer:1973fx} of particle physics, based on the gauge group $SU(3)_C \times SU(2)_L \times U(1)_Y$, has successfully described a wide range of phenomena involving fundamental particles and interactions. However, several observations remain unexplained within its framework, including the origin of neutrino~\cite{Cowan:1956rrn} masses, the nature of their mass ordering, and the hierarchy among charged lepton masses. Notably, neutrino oscillation experiments have firmly established that neutrinos are massive and undergo flavour transitions~\cite{SNO:2002tuh, KamLAND:2002uet, Super-Kamiokande:1998uiq}, necessitating an extension beyond the SM. Among various beyond SM (BSM) approaches, seesaw mechanisms~\cite{Yoshimura:1978ex, Akhmedov:1999tm} provide elegant explanations for the smallness of neutrino masses. In particular, the Type-I seesaw~\cite{Mohapatra:2004zh, King:2003jb} introduces heavy right handed neutrinos\,($\nu_{R}$), while the Type-II seesaw~\cite{King:2003jb, Cheng:1980qt} relies on an $SU(2)_L$ scalar triplet\,($\Delta$) acquiring a small vacuum expectation value\,(vev). The typical Feynman diagrams for the Type-I and Type-II seesaw are shown in Fig.\,\ref{fig:feynman}. An interesting scenario combining both, commonly known as the Type-I+II seesaw mechanism~\cite{Cai:2017mow, Akhmedov:2006de, Verma:2018lro, Singh:2022nmk}, offers increased structural richness and emerges naturally in symmetry driven frameworks. The neutrino mass matrix, $M_\nu$, encodes valuable information about mixing angles\,($\theta_{12}, \theta_{13}, \theta_{23}$), CP violating phases\,($\delta, \alpha, \beta$), and the neutrino mass eigenvalues\,($m_1, m_2, m_3$). However, neutrino oscillation data constrains only a subset of these parameters. To bridge this gap, various theoretical strategies have been explored, most notably the introduction of textures, where specific correlations or patterns in the matrix elements reduce parameter freedom and enhances the predictive power\,\cite{Ramond:1993kv, Fritzsch:2011qv, Ludl:2014axa, Harrison:2002er, Xing:2022uax}. 

	In this work, we propose a flavour motivated model embedded within a Type-I+II seesaw framework, where a discrete symmetry structure gives rise to a distinctive neutrino mass texture. This texture imposes strong constraints on the neutrino mixing parameters. Moreover, the model naturally reproduces the observed charged lepton mass hierarchy through symmetry protected suppressions. We also explore the phenomenological consequences of the texture, particularly its imprints on low energy observables such as neutrinoless double beta\,($0\nu\beta\beta$) decay, charged lepton flavour violation\,(cLFV) and the CP assymetry parameter in neutrino oscillation\,($A_{\mu e}$).

	The structure of the paper is as follows: in Section~\ref{section 2}, we introduce the model and the underlying symmetry groups, and discuss the construction of the model in detail. Section~\ref{section 3} contains the numerical analysis and the predictions of the model. In Sections~\ref{section 4}, \ref{section 5}, and \ref{section 6}, we explore the phenomenological implications of the model by studying its predictions for $0\nu\beta\beta$ decay, cLFV, and $A_{\mu e}$, respectively. Finally, in Section~\ref{section 7}, we present a summary and discussion of our findings.

\section{Theoretical Framework}
\label{section 2}
We extend the Standard Model (SM) gauge group by incorporating an $A_4$ flavour symmetry\,\cite{Ma:2001dn, King:2006np, Altarelli:2010gt, King:2013eh} along with cyclic $Z_N$\,\cite{Ma:2004yx,Hu:2006wk,CarcamoHernandez:2020udg, Dey:2023rht, Dey:2024ctx, Van:2024qrs} symmetries, particularly $Z_3$\,\cite{CarcamoHernandez:2020udg, Dey:2023rht, Dey:2024ctx} and $Z_{10}$\,\cite{CarcamoHernandez:2020udg, Dey:2023rht, Dey:2024ctx}. The complete symmetry group of the model is given by
\begin{equation}
\mathcal{G}= A_4 \times Z_3 \times Z_{10}\times SU(2)_L \times U(1)_Y.
\end{equation}
The SM field content is augmented by introducing $SU(2)_L$ singlet right handed neutrino fields ($\nu_{l_R}$) and five $SU(2)_L$ singlet scalar fields ($\chi, \kappa, \sigma, \rho, \zeta$). In addition, we introduce an extra $SU(2)_L$ doublet $\Phi$ and a triplet $\Delta$. The cyclic symmetries play a crucial role in restricting potential next to leading order\,(NLO) corrections, which could otherwise perturb the Yukawa Lagrangian ($\mathcal{L}_Y$). The transformation properties of the field content under the full symmetry of the model are summarized in Table\,\ref{tab:1}. 
\begin{table}[h!]
\centering
\small
\setlength{\tabcolsep}{8.2pt}
\renewcommand{\arraystretch}{1}
\begin{tabular}{lcccccccccccc}
\hline
 Field & $D_{l_L}$ & $l_R$ & $H$ & $\nu_{l_R}$ & $\Phi$ & $\Delta$ & $\chi$ & $\kappa$ & $\sigma$ & $\rho$ & $\zeta$ \\
\hline
$A_4$        & 3 & (1,\,1$''$,\,1$'$) & 3 & (1,\,1$''$,\,1$'$) & 3 & 3 & 1$'$ & 1$''$ & 1 & 1 & 1 \\
$Z_3$        & $\omega$ & $\omega$ & 1 & (1, $\omega^*$, $\omega^*$) & $\omega$ & $\omega^*$ & $\omega^*$ & $\omega^*$ & $\omega$ & $\omega$ & $\omega^*$ \\
$Z_{10}$     & 0 & (1, 4, 7) & 0 & (0, 2, 7) & 0 & 0 & 6 & 6 & 8 & 3 & 1 \\
$SU(2)_L$    & 2 & 1 & 2 & 1 & 2 & 3 & 1 & 1 & 1 & 1 & 1 \\
$U(1)_Y$     & -1 & -2 & 1 & 0 & -1 & -2 & 0 & 0 & 0 & 0 & 0 \\
\hline
\end{tabular}
\caption{Transformation properties of the fields under $\mathcal{G}$.}
\label{tab:1}
\end{table}
The leading order\,(LO) $\mathcal{L}_Y$ is constructed in the Altarelli–Feruglio basis\,\cite{Altarelli:2005yx} of $A_4$, and is presented as shown below
\begin{eqnarray}
- \mathcal{L}_Y &=& \mathcal{L}_{l}+ \mathcal{L}_{\nu}+ h.c.
\end{eqnarray}
Here, $\mathcal{L}_{l}$ and $\mathcal{L}_{\nu}$ correspond to the charged lepton and neutrino sectors, respectively, and are structured as follows,
\begin{equation}
\mathcal{L}_{l} = y_e\, \bar{D}_{l_L} \,H\, e_{R} \left(\frac{\zeta}{\Lambda}\right)^9 +\, y_{\mu}\, \bar{D}_{l_L} \,H\, \mu_{R} \left(\frac{\zeta}{\Lambda}\right)^6 +\, y_{\tau}\, \bar{D}_{l_L} \,H\, \tau_{R} \left(\frac{\zeta}{\Lambda}\right)^3,
\label{clagrangian}
\end{equation}
and
\begin{eqnarray}
\mathcal{L}_{\nu} &=& y_{D_1} \bar{D}_{l_L}\Phi\, \nu_{e_R} + y_{D_2} \bar{D}_{l_L}\Phi\,\nu_{\mu_R} \left(\frac{\sigma}{\Lambda}\right) + y_{D_3} \bar{D}_{l_L}\Phi\,\nu_{\tau_R} \left(\frac{\rho}{\Lambda}\right) + \frac{1}{2} \overline{\nu^C_{e_R}} \nu_{e_R} + \frac{1}{2} y_1 \overline{\nu^C_{\mu_R}} \nu_{\tau_R}\zeta  + \frac{1}{2} y_1  \overline{\nu^C_{\tau_R}} \nu_{\mu_R} \zeta + \frac{1}{2}y_2 \overline{\nu^C_{\mu_R}} \nu_{\mu_R} \kappa \,\nonumber\\&& + \frac{1}{2}y_3 \overline{\nu^C_{\tau_R}} \nu_{\tau_R} \chi + \frac{1}{2} y_{\Delta} \bar{D}_{l_L} D^C_{l_L} i \sigma_2 \Delta,
\label{nLagrangian}
\end{eqnarray} 
where $\Lambda$ is the effective scale of the theory. The symmetry of $\mathcal{L}_Y$ is broken down when the scalar fields acquire their respective vevs.
For the following vacuum alignment: $\langle H \rangle = v_h(1,0,0)^T$, $\langle \Phi \rangle = v_{\Phi}(0,0,1)^T$, $\langle \Delta \rangle = v_{\Delta}(-1,0,-1)^T$, $\langle \chi \rangle = v_{\chi}$, $\langle \kappa \rangle = v_{\kappa}$, $\langle \sigma \rangle = v_{\sigma}$, $\langle \rho \rangle = v_{\rho}$, $\langle \zeta \rangle = v_{\zeta}$, one can derive the structure of the fermion mass matrices. Dealing with the vev alignment in the context of neutrino model building is often overlooked. However, in the present scenario, a detailed discussion on the scalar potential is shown in the Appendix\,\ref{Appendix}. 

	With the vacuum alignment in place, we now proceed to discuss the charged lepton mass hierarchy. The model predicts a diagonal charged lepton mass matrix, with masses satisfying the hierarchical relation
\begin{equation}
m_e : m_{\mu} : m_{\tau} = \lambda_f^6\, y_e : \lambda_f^3\, y_{\mu} : y_{\tau},
\end{equation}
where $\lambda_f = v_{\zeta}/\Lambda$. This relation can successfully account for the observed hierarchy\,\cite{ParticleDataGroup:2018ovx, ParticleDataGroup:2022pth} among the charged lepton masses.

	The complex symmetric Majorana neutrino mass matrix in the Type-I+II seesaw framework is given by $M_{\nu} = M_{II} - M_D M_R^{-1} M_D^T$, where $M_D$, $M_R$, and $M_{II}$ denote the Dirac neutrino mass matrix, the right handed Majorana neutrino mass matrix, and the mass matrix arising from the Type-II seesaw mechanism, respectively.
In the present model, the effective Type-I+II neutrino mass matrix takes the following form
\begin{equation}
M_{\nu} =
\begin{pmatrix}
r - 2t & t & -q \\
t & -p & t \\
-q & t & s - 2t
\end{pmatrix},
\label{neutrinomassmatrix}
\end{equation}
where
\begin{eqnarray}
\label{eqmt1}
p &=& \frac{y^2_{D_1} v^2_{\Phi}}{M_1},\\
q &=& \frac{y_{D_2}y_{D_3} y_1 v^2_{\Phi} v_{\rho} v_{\sigma} v_{\zeta}}{\left((y_1 v_{\zeta})^2- y_2 y_3 v_{\kappa} v_{\chi}\right) \Lambda^2},\\
r &=& \frac{y^2_{D_3} y_2 v^2_{\Phi} v^2_{\rho} v_{\kappa}}{\left((y_1 v_{\zeta})^2- y_2 y_3 v_{\kappa} v_{\chi}\right) \Lambda^2},\\
s &=& \frac{y^2_{D_2} y_3 v^2_{\Phi} v^2_{\sigma} v_{\chi}}{\left((y_1 v_{\zeta})^2- y_2 y_3 v_{\kappa} v_{\chi}\right) \Lambda^2},\\
t &=& y_{\Delta} v_{\Delta}.
\label{eqmt2}
\end{eqnarray}
For the sake of simplicity, we refer to the parameters on the right hand side of Eqs.\,(\ref{eqmt1})–(\ref{eqmt2}) as model parameters, and those on the left hand side as texture parameters.

The texture in Eq.\,(\ref{neutrinomassmatrix}) features a unique relation
\begin{equation}
(M_{\nu})_{e\mu} = (M_{\nu})_{\mu\tau}.
\end{equation}
We refer to this type of texture as the \emph{Echo 12–23 Texture} because the (12) and (23) elements of $M_\nu$ echo each other through their equality. This correlation may significantly restrict the allowed values of the observable parameters. Before delving into the predictive implications of this texture, we first briefly review the parametrization of the PMNS matrix ~\cite{Maki:1962mu}.
	According to the Particle Data Group (PDG)~\cite{ParticleDataGroup:2018ovx}, the Pontecorvo Maki Nakagawa Sakata\,(PMNS) matrix\,($U$) is parametrized in the following way
\begin{equation}
\label{pmns}
U = P_{\phi} \, \tilde{U} \, P_M,
\end{equation}
where $P_{\phi} = \mathrm{diag}(e^{i\phi_1}, e^{i\phi_2}, e^{i\phi_3})$ contains the unphysical phases\,($\phi_1, \phi_2, \phi_3$), and $ P_M = \mathrm{diag}(e^{i\alpha}, e^{i\beta}, 1)$ encodes the Majorana phases. 
The arbitrary phases $\phi_i$ ($i = 1,2,3$) are absorbed by redefining the charged lepton fields, simplifying the PMNS matrix to
\begin{equation}
U = \tilde{U} P_M.
\label{pmns1}
\end{equation}
In the charged lepton diagonal basis, $U$ diagonalizes $M_\nu$ as
\begin{equation}
U^T M_\nu U = \mathrm{diag}(m_1, m_2, m_3).
\label{11}
\end{equation}
This leads to the equivalent condition
\begin{equation}
\tilde{U}^T M_\nu \tilde{U} = \mathrm{diag}(\tilde{m}_1, \tilde{m}_2, m_3),
\label{13}
\end{equation}
where the rephased masses are defined as $\tilde{m}_1 = m_1 e^{-2i\alpha}$ and $\tilde{m}_2 = m_2 e^{-2i\beta} $. Thus, $\tilde{U}$ is the matrix that diagonalizes  the $M_{\nu}$ of our model. The general form of $\tilde{U}$ is shown below
\begin{equation}
\tilde{U}=
\begin{pmatrix}
c_{12} c_{13} & s_{12} c_{13} & s_{13} e^{-i\delta} \\
- s_{12} c_{23} - c_{12} s_{13} s_{23} e^{i\delta} & c_{12} c_{23} - s_{12} s_{13} s_{23} e^{i\delta} & c_{13} s_{23} \\
s_{12} s_{23} - c_{12} s_{13} c_{23} e^{i\delta} & - c_{12} s_{23} - s_{12} s_{13} c_{23} e^{i\delta} & c_{13} c_{23}
\end{pmatrix},
\end{equation}
where $c_{ij}= \cos \theta_{ij}$ and $s_{ij}= \sin \theta_{ij}$. The $\theta_{ij}$ represents the three mixing angles. 

	The diagonalizing matrix $\tilde{U}$ imposes constraints on the neutrino mass matrix $M_{\nu}$, such that all complex texture parameters $p$, $q$, $r$, $s$, and $t$ can be expressed in terms of only three real parameters: $Re[s], Im[s]$ and $Re[t]$. In other words, these three free parameters are sufficient to fully determine $M_{\nu}$ and its associated predictions.

\section{Numerical Analysis}
\label{section 3}

We first perform a numerical analysis in the light of normal hierarchy\,(NH) of neutrino masses. In this regard, the $3\sigma$ ranges of the mixing angles and the Dirac CP phase $\delta$ are used as inputs. A large sample of points in the parameter space ${Re[s], Im[s], Re[t]}$ is generated, requiring that the mass-squared differences $\Delta m_{21}^2$ and $\Delta m_{31}^2$ lie within the $3\sigma$ ranges reported by NuFIT~6.0\,\cite{Esteban:2024eli}. In addition, the sum of the three light neutrino masses is required to satisfy the cosmological bound $\sum m_i < 0.12\,\text{eV}$\,\cite{Planck:2018vyg}. From the dataset, we find that viable parameter points exist within the following ranges
\begin{gather*}
(Re[s])_{\text{min}} = 0.0752\,\text{eV}, \quad (Re[s])_{\text{max}} = 0.0843\,\text{eV};\\
(Im[s])_{\text{min}} = -0.0089\,\text{eV}, \quad (Im[s])_{\text{max}} = 0.0086\,\text{eV};\\
(Re[t])_{\text{min}} = 0.0192\,\text{eV}, \quad (Re[t])_{\text{max}} = 0.0236\,\text{eV},
\end{gather*}

However, when performing a similar scan assuming an inverted hierarchy\,(IH), we find that the model yields no viable parameter points. \emph{Thus, IH is disfavoured in this model}.

It is important to note that, although the mixing angles and $\delta$ are scanned over their full $3\sigma$ ranges, we observe that for the viable parameter space yielding $\sum m_i$ within the range 0.1 eV to 0.118 eV, the Dirac CP phase $\delta$ is strongly restricted. A large portion of its allowed range, approximately between $96^\circ$ and $285.6^\circ$, is forbidden by the model. While $\delta$ is typically treated as a free parameter in many flavour models, here its viable values are tightly limited due to the structure of the \emph{Echo 12–23 Texture}. As a result, the overall parameter space through this restriction on $\delta$ is also constrained. From the predicted spectra of the neutrino mass eigenvalues, $m_1$ and $m_2$ appears to overlap in certain regions, potentially giving the impression of an exact degeneracy. However, when we calculate $m_2 - m_1$, the difference is never zero, thereby removing the notion of exact degeneracy. This supports a quasi-degenerate mass spectra, i.e ., $m_1 \lesssim m_2 \lesssim m_3$. The model predicts the Majorana phases within a narrow range. The phases $\alpha$ and $\beta$ are constrained to lie between $-90^{\circ}$ and $90^{\circ}$ and $-12.7^{\circ}$ and $13.6^{\circ}$, respectively. Since the Dirac CP phase $\delta$ is subject to restrictions, we also study the associated Jarlskog invariant $J_{\text{CP}}$, which quantifies leptonic CP violation and depends directly on $\delta$ and the mixing angles.
\begin{figure}
    \centering
    % First row: three figures
    \begin{minipage}[b]{0.305\textwidth}
        \centering
        \includegraphics[width=\textwidth]{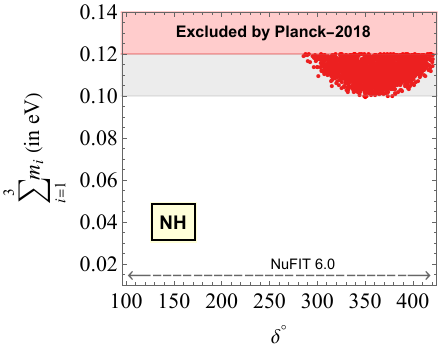}
        (a)
        \label{fig:1a}
    \end{minipage}
    \hfill
    \begin{minipage}[b]{0.3\textwidth}
        \centering
        \includegraphics[width=\textwidth]{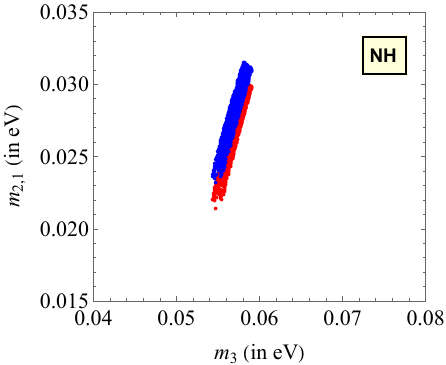}
        (b)
        \label{fig:1b}
    \end{minipage}
    \hfill
    \begin{minipage}[b]{0.3\textwidth}
        \centering
        \includegraphics[width=\textwidth]{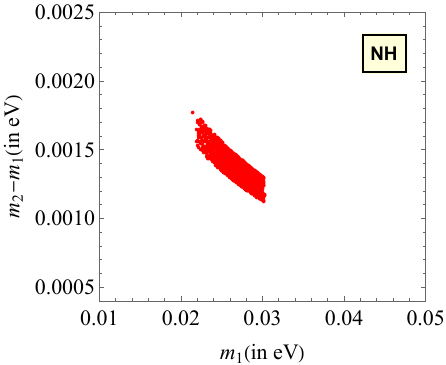}
       (c)
        \label{fig:1c}
    \end{minipage}
    \begin{minipage}[b]{0.29\textwidth}
        \centering
        \includegraphics[width=\textwidth]{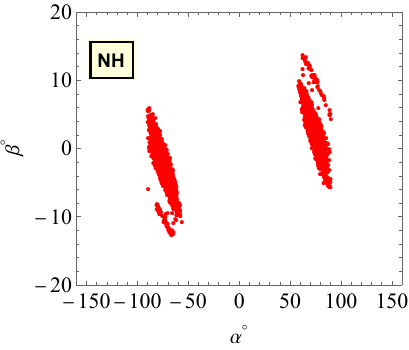}
        (d)
        \label{fig:1e}
    \end{minipage}
    \hspace{0.25cm}
    \begin{minipage}[b]{0.29\textwidth}
        \centering
        \includegraphics[width=\textwidth]{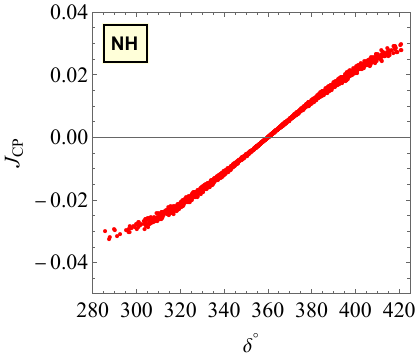}
       (e)
        \label{fig:1f}
    \end{minipage}
    \caption{The plots of the predicted parameters are shown. (a) $\delta$ vs $ \sum m_i$. (b) $m_{2,1}$ vs $m_3$. (c) $m_2-m_1$ vs $m_1$. (d) $\alpha$ vs $\beta$. (e) $\delta$ vs $J_{\text{CP}}$.}
    \label{fig:1}
\end{figure}
The $J_{\text{CP}}$ is predicted to lie between $-0.0324$ and $0.030$. To summarize our findings, we list the numerical values of the predicted parameters in  Table\,\ref{tab:2}. To have a graphical understanding of the predicted parameters, we highlight the plots for the same in Fig.\,\ref{fig:1}. In Fig.\,\ref{fig:2}, we highlight the plots of $Re[s], Im[s], Re[t]$.
\begin{table}[h!]
\centering
\setlength{\tabcolsep}{4.2pt}
\renewcommand{\arraystretch}{1.5}
\begin{tabular}{lcccccccc}
\hline
 Prediction & $\delta$ & $m_1$ & $m_2$ & $m_3$ & $\sum m_i$ & $\alpha$ & $\beta$ & $J_{\text{CP}}$ \\
\hline
Minimum & $285.7^{\circ}$ & 0.021\,eV & 0.023\,eV & 0.053\,eV & 0.1\,eV & $-90^{\circ}$ & $-12.7^{\circ}$ & $-0.0324$ \\
Maximum & $421^{\circ}$ & 0.030\,eV & 0.031\,eV & 0.058\,eV & 0.118\,eV & $90^{\circ}$ & $13.6^{\circ}$ & $0.030$ \\
\hline
\end{tabular}
\caption{The approximate minimum and maximum values of $\delta$, $m_1$, $m_2$, $m_3$, $\sum m_i$, $\alpha$, $\beta$, and $J_{\text{CP}}$ as predicted by the model.}
\label{tab:2}
\end{table}
So far, we have presented the predictions for the low energy observable quantities as well as the texture parameters $Re[s]$, $Im[s]$, and $Re[t]$. These texture parameters, however, are not arbitrary, they are generated from a deeper set of model parameters, including the Yukawa couplings and flavon vevs, as outlined in Eqs.\,(\ref{eqmt1})–(\ref{eqmt2}). Therefore, by using the predicted values of the texture parameters, we can determine or constrain the underlying model parameters that define the theory at a more fundamental level. Though the framework cannot directly predict the individual model parameters, it can provide numerical information for several interesting combinations of them.
\begin{figure}[h!]
    \centering
    \begin{minipage}[b]{0.29\textwidth}
        \centering
        \includegraphics[width=\textwidth]{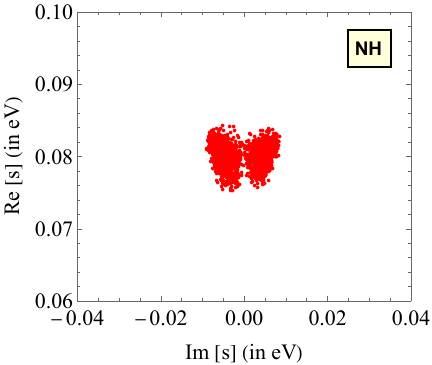}
        (a)
        \label{fig:2a}
    \end{minipage}
    \hspace*{0.3cm}
    \begin{minipage}[b]{0.29\textwidth}
        \centering
        \includegraphics[width=\textwidth]{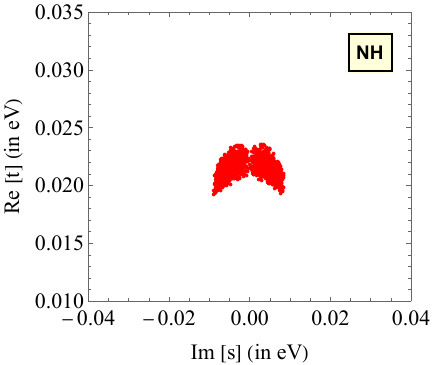}
        (b)
        \label{fig:2b}
    \end{minipage}
    \caption{The plots of the free parameters are shown. (a)$Im[s]$ vs $Re[s]$. (b) $Im[s]$ vs $Re[t]$.}
\label{fig:2}
\end{figure}
In this connection, we highlight the model parameter combinations below, which are functions of the free parameters $Re[s]$, $Im[s]$, and $Re[t]$
\begin{align}
& G_1 = \frac{y_{D_1} v_{\Phi}}{\sqrt{M_1}}, \quad
G_2 = \frac{y_{D_2} v_{\Phi} v_{\sigma}}{\sqrt{y_2 v_{\kappa}} \Lambda}, \quad
G_3 = \frac{y_{D_3} v_{\Phi} v_{\rho}}{\sqrt{y_3 v_{\chi}} \Lambda}, \nonumber\\
& \hspace{1.5cm} G_4 = \frac{y_1 v_{\zeta}}{\sqrt{y_2 y_3 v_{\kappa} v_{\chi}}}, \quad
G_5 = y_{\Delta} v_{\Delta}.
\end{align}
The predicted numerical values of the model parameter combinations $G_1, G_2, G_3, G_4$, and $G_5$ are listed in Table\,\ref{tab:3}.
\begin{table}[h!]
\centering
\setlength{\tabcolsep}{6pt}
\renewcommand{\arraystretch}{1.2}
\begin{tabular}{lcc}
\hline
Parameter & Minimum & Maximum \\
\hline
$|G_1|$ & 0.156 $\text{eV}^{\frac{1}{2}}$ & 0.253 $\text{eV}^{\frac{1}{2}}$ \\
$|G_2|$ & 0.05 $\text{eV}^{\frac{1}{2}}$ & 0.355 $\text{eV}^{\frac{1}{2}}$ \\
$|G_3|$ & 0.025 $\text{eV}^{\frac{1}{2}}$ & 0.224 $\text{eV}^{\frac{1}{2}}$ \\
$|G_4|$ & 0.118 & 1.185 \\
$|G_5|$ & 0.060 eV & 0.0722 eV \\
$Arg[G_1]$ & $-90.000^{\circ}$ & $90.000^{\circ}$ \\
$Arg[G_2]$ & $-92.364^{\circ}$ & $92.233^{\circ}$ \\
$Arg[G_3]$ & $-92.854^{\circ}$ & $92.767^{\circ}$ \\
$Arg[G_4]$ & $-42.100^{\circ}$ & $50.000^{\circ}$ \\
$Arg[G_5]$ & $-21.170^{\circ}$ & $23.000^{\circ}$ \\
\hline
\end{tabular}
\caption{The approximate maximum and minimum values of the model parameter combinations (transposed).}
\label{tab:3}
\end{table}
To visualise these combinations graphically, we plot them in Fig.\,\ref{fig:3}.
\begin{figure}[h!]
    \centering
    % First row: three figures
    \begin{minipage}[b]{0.272\textwidth}
        \centering
        \includegraphics[width=\textwidth]{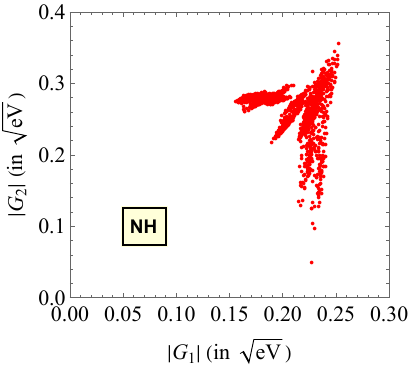}
        (a)
        \label{fig:3a}
    \end{minipage}
    \hfill
    \begin{minipage}[b]{0.273\textwidth}
        \centering
        \includegraphics[width=\textwidth]{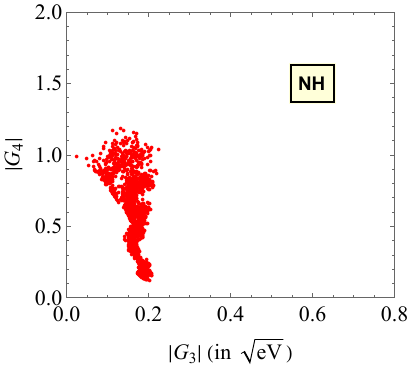}
        (b)
        \label{fig:3b}
    \end{minipage}
    \hfill
    \begin{minipage}[b]{0.28\textwidth}
        \centering
        \includegraphics[width=\textwidth]{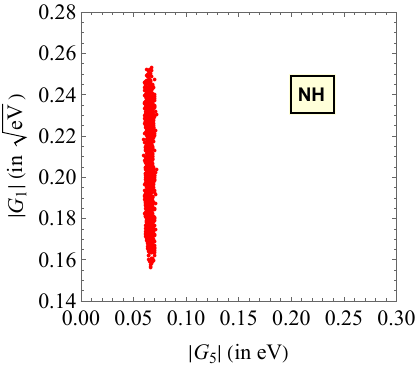}
       (c)
        \label{fig:3c}
    \end{minipage}
   \hfill
    \begin{minipage}[b]{0.28\textwidth}
        \centering
        \includegraphics[width=\textwidth]{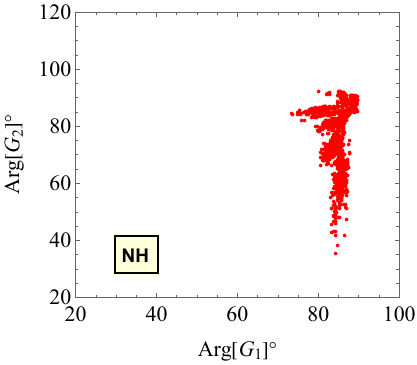} % replace with your figure
        (d)
        \label{fig:3d}
    \end{minipage}
    \hfill
    \begin{minipage}[b]{0.28\textwidth}
        \centering
        \includegraphics[width=\textwidth]{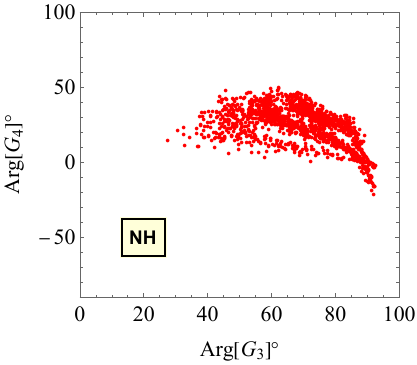} % replace with your figure
        (e)
        \label{fig:3e}
    \end{minipage}
    \hfill
    \begin{minipage}[b]{0.28\textwidth}
        \centering
        \includegraphics[width=\textwidth]{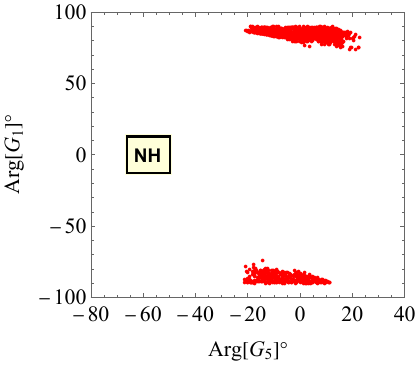} % replace with your figure
        (f)
        \label{fig:3f}
    \end{minipage}
    \caption{The plots of the model parameter combinations are shown.}
    \label{fig:3}
\end{figure}

	The model is now equipped with all the necessary numerical inputs to examine its consistency in light of low energy phenomenologies such as $0\nu\beta\beta$ decay, cLFV and $A_{\mu e}$.
\section{Neutrinoless Double Beta Decay}
\label{section 4}
\begin{figure}[h]
    \centering
    % First row: three figures
    \begin{minipage}[b]{0.28\textwidth}
        \centering
        \includegraphics[width=\textwidth]{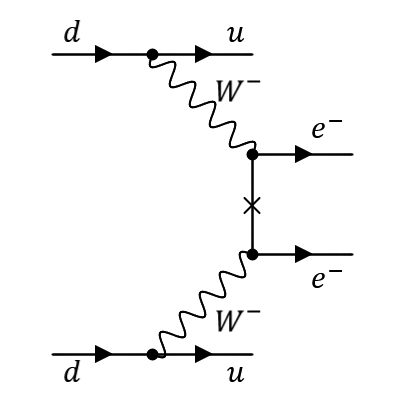}
    \end{minipage}
    \caption{Feynman diagram for $0\nu\beta\beta$ decay.}
    \label{fig:nbbdecay}
\end{figure}

The model adheres to the Majorana nature of the neutrinos, therefore, it is pertinent to discuss the effective Majorana mass, $m_{\beta\beta}$. This quantity is a crucial observable in determining the  half life\,($T_{1/2}$) of $0 \nu \beta \beta$ decay \cite{Schechter:1981bd, Blennow:2010th, Barabash:2023dwc}. The $0 \nu \beta \beta$ decay is a hypothetical decay  in which an unstable nucleus undergoes two simultaneous beta decays without emitting neutrinos. Such a decay is only possible if neutrinos are Majorana particles. A typical Feynman diagram for this process is shown in Fig.\,\ref{fig:nbbdecay}.
A confirmed observation of this process would thus provide direct evidence for the Majorana nature of neutrinos. Although no positive signal has been observed to date, several experiments have placed stringent upper limits on the observable $m_{\beta\beta}$, which are summarised in Table\,\ref{tab:4}. The general expression for
$m_{\beta\beta}$ is given below
\begin{equation}
m_{\beta\beta} = |U^2_{ei} m_i|, \quad \text{where,}\,\, i=1,2,3.
\end{equation}
\begin{table}[h]
\centering
\setlength{\tabcolsep}{5pt}
\renewcommand{\arraystretch}{1.2}
\begin{tabular}{lcc}
\hline
Isotope & $m_{\beta\beta}$ (eV) & Collaboration \\
\hline
${}^{130}\mathrm{Te}$ & $< 0.09$–$0.305$ & CUORE\,\cite{CUORE:2021mvw} \\
${}^{136}\mathrm{Xe}$ & $< 0.093$–$0.286$ & EXO\,\cite{EXO-200:2019rkq} \\
${}^{76}\mathrm{Ge}$ & $< 0.08$–$0.18$ & GERDA\,\cite{GERDA:2020xhi} \\
${}^{136}\mathrm{Xe}$ & $< 0.061\text{–}0.165,\, < 0.036\text{–}0.156$ & KamLAND-Zen\,\cite{KamLAND-Zen:2016pfg,KamLAND-Zen:2022tow} \\
${}^{76}\mathrm{Ge}$ & $< 0.009$–$0.021$ & LEGEND-1000\,\cite{LEGEND:2021bnm} \\
\hline
\end{tabular}
\caption{The present upper limits on $m_{\beta\beta}$ from $0\nu\beta\beta$ decay searches in different isotopes, along with the expected sensitivity of the future LEGEND-1000 experiment.}
\label{tab:4}
\end{table}
\begin{figure}[h!]
    \centering
    % First row: three figures
    \begin{minipage}[b]{0.38\textwidth}
        \centering
        \includegraphics[width=\textwidth]{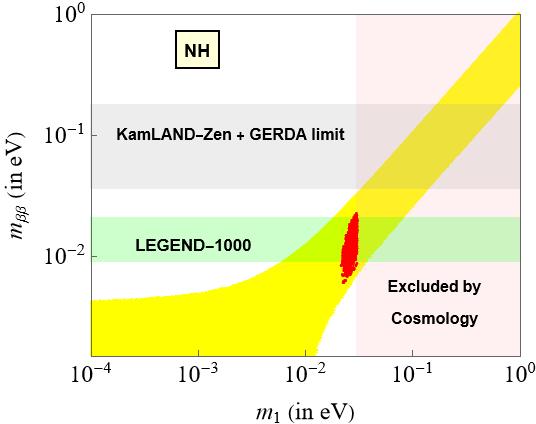}
    \end{minipage}
    \caption{The plot of $m_{\beta \beta}$ vs $m_1$. In the plot, the `yellow' band shows the generic region allowed by current oscillation data in case of NH. The `gray' region indicates the bound on $m_{\beta\beta}$ set by KamLAND-Zen and GERDA. The `faded pink' region corresponds to the cosmologically excluded region for NH. The `light green' region shows the projected sensitivity of the future LEGEND-1000 experiment, and the `red' regions represent the predictions of the present model.}
    \label{fig:mbb}
\end{figure}
As mentioned previously, $Re[s]$, $Im[s]$, and $Re[t]$ dictate the structure of $M_{\nu}$ and all associated predictions. Hence, $m_{\beta\beta}$ is no exception. For NH, the model places stringent bound on $m_{\beta\beta}$, predicting it to lie within the range 0.0060 eV – 0.0227 eV. This range is consistent with the current upper limits set by KamLAND-Zen and GERDA, and lies within the projected sensitivity of the future LEGEND-1000 experiment. To visualise the prediction of $m_{\beta\beta}$ by the model, we show the plot of $m_{\beta\beta}$ versus the lightest neutrino mass eigenvalue\,($m_1$) in Fig.\,\ref{fig:mbb}
\section{Charged Lepton Flavour Violation}
\label{section 5}
In the SM of particle physics, cLFV processes\,\cite{Calibbi:2017uvl,Ardu:2022sbt,Cei:2014jtm,Matsuzaki:2008ik,Akeroyd:2009nu,PhysRevD.89.013008,Hue:2013uw,Dinh:2013vya,Omura:2015xcg,Zhou:2016ynv,PhysRevD.99.035020,Hernandez-Tome:2018fbq,Calcuttawala:2018wgo,Ferreira:2019qpf} are highly suppressed, with predicted branching ratios around $10^{-50}$. Therefore, cLFV is one of the most promising probes of physics beyond the SM.
Among the various decay channels, transitions involving muons are the most interesting, as muons have a longer lifetime than other leptons and are abundantly produced in cosmic radiation. This makes muon decays ideal for study. So far, there is no experimental evidence of cLFV decays. However, ongoing experiments are steadily improving their sensitivities to search for these rare processes. To complement these experimental efforts, it is worthwhile to explore theoretical frameworks that could explain or predict such decays. As an extension of our model, we plan to study the dominant decay $\mu \rightarrow e\gamma$.
The current best limit on the branching ratio of the $\mu \rightarrow e\gamma$ decay is provided by MEG\,\cite{MEG:2016leq}
\begin{equation}
\text{BR}(\mu \rightarrow e\gamma) < 4.2 \times 10^{-13}.
\end{equation}
The MEG II\,\cite{Meucci:2022qbh,MEGII:2023ltw,Francesconi:2024bpo} experiment is currently searching for this rare decay at the Paul Scherrer Institute (PSI) muon beam facility. Its goal is to improve the sensitivity of measurements to $BR(\mu \rightarrow e\gamma) \sim 6 \times 10^{-14}$ by the end of 2026. In the present model, it is possible to explore how each seesaw mechanism may contribute to the $\mu \rightarrow e\gamma$ decay. We then aim to determine which contribution is the dominant one.
In the case of Type-I seesaw, the $\mu \rightarrow e\gamma$ decay arises due to the existence of heavy-light neutrino mixing. To understand this, we write the effective Lagrangian for the Type-I seesaw in the following manner
\begin{eqnarray}
\mathcal{L}_{\nu}&=& \frac{1}{2}\overline{n^c} M n + h.c,
\label{NNLagrangian}
\end{eqnarray}
with 
\begin{eqnarray}
n &=& \begin{pmatrix}
\nu^c_L\\
\nu_R\\
\end{pmatrix},
\end{eqnarray}
and the $6\times6$ mass matrix,
\begin{eqnarray}
M &=& \begin{pmatrix}
0 & M_D\\
M^T_{D}   & M_R\\
\end{pmatrix}.
\end{eqnarray}
The block diagonalisation of $M$ is achieved by the following transformation
\begin{equation}
\begin{pmatrix}
1-\frac{R R^{\dagger}}{2} & R\\
-R^{\dagger}   & 1-\frac{R R^{\dagger}}{2}\\
\end{pmatrix}^T 
\begin{pmatrix}
0 & M_D\\
M^T_{D}   & M_R\\
\end{pmatrix}
\begin{pmatrix}
1-\frac{R R^{\dagger}}{2} & R\\
-R^{\dagger}  & 1-\frac{R R^{\dagger}}{2}\\
\end{pmatrix} \simeq 
\begin{pmatrix}
m_{\text{light}} & 0\\
0 & M_{\text{heavy}}\\
\end{pmatrix}, 
\label{blockdiagonalisation}
\end{equation}
where $R = M^{\dagger}_D M_R^{-1}$, $m_{\text{light}}= - M_D M_R^{-1} M_D^T$ and $M_{\text{heavy}}= M_R$. The next step is to diagonalise the block diagonalised matrix in Eq.\,\ref{blockdiagonalisation}. This is achieved via the transformation below
\begin{equation}
\begin{pmatrix}
U & 0\\
0 & U_R\\
\end{pmatrix}^T 
\begin{pmatrix}
m_{\text{light}} & 0\\
0 & M_{\text{heavy}}\\
\end{pmatrix}
\begin{pmatrix}
U & 0\\
0 & U_R\\
\end{pmatrix} \simeq 
\begin{pmatrix}
(m_i)^{\text{diag}} & 0\\
0 & (M_i)^{\text{diag}}\\
\end{pmatrix}= M^{\text{diag}}, 
\label{diagonalistaion}
\end{equation}
where $U$ is the PMNS matrix that diagonalises $m_{\text{light}}$, and $U_R$ diagonalises $M_{\text{heavy}}$. This two step diagonalisation of $M$ can be simplified to a single step in the following way
\begin{equation}
\begin{pmatrix}
U & 0\\
0 & U_R\\
\end{pmatrix}^T \begin{pmatrix}
1-\frac{R R^{\dagger}}{2} & R\\
-R^{\dagger}   & 1-\frac{R R^{\dagger}}{2}\\
\end{pmatrix}^T \begin{pmatrix}
0 & M_D\\
M^T_{D}   & M_R\\
\end{pmatrix} \begin{pmatrix}
1-\frac{R R^{\dagger}}{2} & R\\
-R^{\dagger}   & 1-\frac{R R^{\dagger}}{2}\\
\end{pmatrix} \begin{pmatrix}
U & 0\\
0 & U_R\\
\end{pmatrix} \simeq M^{\text{diag}}.
\end{equation}
Therefore, $M$ can be diagonalised by the matrix
\begin{equation}
D= \begin{pmatrix}
(1-\frac{R R^{\dagger}}{2})U & R U_R\\
-R^{\dagger} U  & (1-\frac{R R^{\dagger}}{2})U_R\\
\end{pmatrix},
\end{equation}
and the off diagonal part $K= R\,U_R = M^{\dagger}_D M_R^{-1} U_R$ carries the information of heavy-light neutrino mixing. In the case of the Type-I seesaw, $K$ directly enters the expression for the branching ratio of the $\mu \rightarrow e \gamma$ decay, which is given by\,\cite{Bilenky:1977du, Ilakovac:1994kj, Tommasini:1995ii, Forero:2011pc, Datta:2021zzf, Dey:2023bfa}
\begin{equation}
BR(\mu \rightarrow e\gamma)= \frac{3 \alpha_f}{8 \pi}\left| \sum_{i} K_{ei}K^{\dagger}_{i\mu} F\left(\frac{M_i^2}{M_W^2}\right) \right|^2,
\end{equation}
where, $\alpha_f$ stands for the fine structure constant, $M_i$ is the mass of the $i^{\text{th}}$ right handed neutrino, and $M_W$ is the mass of $W$ boson. The factor $F(x)$ is expressed as
\begin{equation}
F(x)=\frac{x(1-6x+3x^2+2x^3-6x^2 \ln{x})}{2(1-x)^4},
\end{equation}
where, $x=\frac{M_i^2}{M_W^2}$.
To compute $BR(\mu \rightarrow e\gamma)$, we consider $M_R$ in the TeV range, requiring $M_D \sim$ MeV to reproduce light neutrino masses $\sim 10^{-2}$ eV. This leads to $BR(\mu \rightarrow e\gamma) \sim \mathcal{O}(10^{-40})$, far below the experimental limit. Hence, the Type-I contribution to $BR(\mu \rightarrow e\gamma)$ is insignificant.

	With this inference in mind, we now turn our attention to the Type-II seesaw contribution to $BR(\mu \rightarrow e\gamma)$. In this case, the process arises at the one loop level, mediated by the singly and doubly charged components of the scalar triplet $\Delta$. The corresponding branching ratio is given by\,\cite{Dinh:2012bp, Ferreira:2019qpf, Barrie:2022ake}
\begin{equation}
BR(\mu \rightarrow e\gamma)= \frac{\alpha_f \left| (Y^{*}_{II} Y_{II})_{e\mu} \right|^2}{192 \pi G_F^2} \left(\frac{1}{m^2_{\Delta^{\pm}}}+ \frac{8}{m^2_{\Delta^{\pm\pm}}}\right)^2,
\label{T2BR}
\end{equation}
where $Y_{II}$ is the Type-II Yukawa matrix, and $G_F$ denotes the Fermi constant. In our analysis, we assume that the charged components of the scalar triplet are approximately degenerate in mass, i.e., $m_{\Delta^{++}} \simeq m_{\Delta^{+}} \simeq m_{\Delta}$. This assumption is well justified in scenarios where the mass splitting among triplet components is small. Substituting the model predicted value of $(Y^{*}_{II} Y_{II})_{e\mu}$ into Eq.\,(\ref{T2BR}), the branching ratio simplifies to
\begin{equation}
BR(\mu \rightarrow e\gamma)\simeq 0.0208 \, \frac{\alpha_f}{\pi G_F^2} \left( \frac{\left|G_5\right|}{m_{\Delta} v_{\Delta}} \right)^4.
\label{T2BRM}
\end{equation}
From Eq.\,(\ref{T2BRM}), it is evident that the branching ratio depends on the parameters $|G_5|$, $m_{\Delta}$, and $v_{\Delta}$. Using the numerical values of $|G_5|$ obtained from our model, we perform a scan over $m_{\Delta}$ and $v_{\Delta}$ to identify the regions of parameter space that yield branching ratios consistent with experimental limits. This allows us to determine the most favourable regions in the $(m_{\Delta}, v_{\Delta}, |G_5|)$ parameter space. Based on our numerical analysis, we find that the predicted branching ratio $BR(\mu \rightarrow e\gamma)$ lies within the range $[10^{-15}, 4 \times 10^{-13}]$, which is compatible with the current MEG bound and the projected sensitivity of MEG-II. While $|G_5|$ remains fixed within its allowed range $[0.06, 0.072]$ eV, three distinct and viable sets of bounds for $m_{\Delta}$ and $v_{\Delta}$ emerge, as summarised in Table\,\ref{tab:BR}.
\begin{table}[h!]
\centering
\setlength{\tabcolsep}{21pt}
\renewcommand{\arraystretch}{1.7}
\begin{tabular}{ccc|cc}
\hline
\multicolumn{5}{c}{$|G_5| \in [0.06, 0.072]$ eV \quad\quad $BR(\mu \rightarrow e\gamma) \in [10^{-15}, 4 \times 10^{-13}]$} \\
\hline
& \multicolumn{2}{c|}{$m_{\Delta}$} & \multicolumn{2}{c}{$v_{\Delta}$} \\
\cline{1-5}
Set & Minimum & Maximum & Minimum & Maximum \\
\hline
A & 870 GeV & 4 TeV & 2.30 eV & 2.50 eV \\
B & 1.00 TeV & 4.50 TeV & 1.98 eV & 2.10 eV \\
C & 1.20 TeV & 8.70 TeV & 1.00 eV & 1.70 eV \\
\hline
\end{tabular}
\caption{Numerical sets of $(m_{\Delta}, v_{\Delta})$ yielding $BR(\mu \rightarrow e\gamma) \in [10^{-15}, 4 \times 10^{-13}]$ for $|G_5| \in [0.06, 0.072]$ eV.}
\label{tab:BR}
\end{table}
\begin{figure}[h]
    \centering
    % First row: three figures
    \begin{minipage}[b]{0.45\textwidth}
        \centering
        \includegraphics[width=\textwidth]{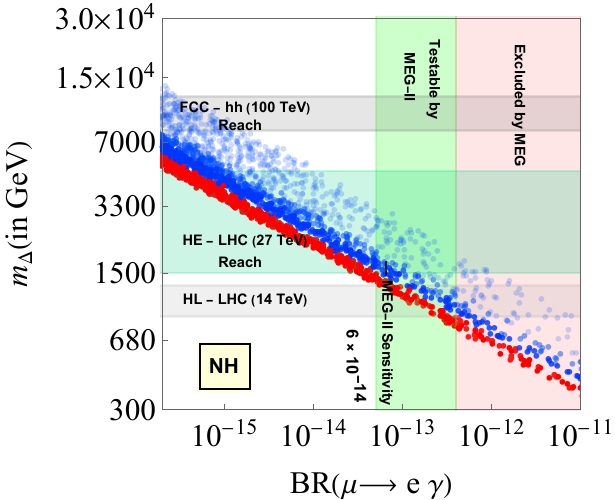}
    \end{minipage}
    \caption{The plot of $BR(\mu \rightarrow e\gamma)$ vs $m_{\Delta}$. The `red', `blue', and `faded blue' points correspond to set A, set B, and set C, respectively. The `faded red' region is excluded by the MEG experiment, while the `pale green' region indicates the $BR(\mu \rightarrow e\gamma)$ values that are within reach of MEG-II sensitivity. The potential reach of high-energy colliders such as HL-LHC, HE-LHC, and FCC-hh is also overlaid for comparison.}
\label{fig:BRnh}
\end{figure}
The HL-LHC\,\cite{Apollinari:2015bam}, HE-LHC\,\cite{Benedikt:2018ofy} and FCC-hh\,\cite{FCC:2018vvp} are expected to probe the higher mass range of the scalar triplet $\Delta$. In particular, the collider signatures of $\Delta^{\pm\pm}$\,\cite{Antusch:2018svb}, such as same-sign dilepton decays ($\Delta^{\pm\pm} \to \ell^\pm \ell^\pm$) for small $v_\Delta$, and diboson decays ($\Delta^{\pm\pm} \to W^\pm W^\pm$) for larger $v_\Delta$, provide promising avenues to test the scalar triplet mass. These searches, combined with charged lepton flavour violation, can play a crucial role in probing the parameter space of our model. To visualize our findings, we present a plot of $BR(\mu \rightarrow e\gamma)$ as a function of $m_{\Delta}$ in Fig.\,\ref{fig:BRnh}.
	Therefore, we may infer that in our model the dominant contribution to the $BR(\mu \rightarrow e\gamma)$ comes from the Type-II seesaw mechanism.
\section{CP Asymmetry in Neutrino Oscillation}
\label{section 6}
The CP asymmetry in neutrino oscillation serves as a crucial observable for testing leptonic CP violation, and is quantified in terms of the Dirac CP-violating phase $\delta$. It appears in the CP asymmetry parameter $A_{\alpha\beta}$, defined as
\begin{equation}
A_{\alpha\beta} = \frac{P(\nu_\alpha \rightarrow \nu_\beta) - P(\bar{\nu}_\alpha \rightarrow \bar{\nu}_\beta)}{P(\nu_\alpha \rightarrow \nu_\beta) + P(\bar{\nu}_\alpha \rightarrow \bar{\nu}_\beta)},
\end{equation}
where $\alpha, \beta = e, \mu, \tau$ and $\alpha \neq \beta$, and $P$ denotes the transition probability. For oscillations between muon and electron flavours, the relevant CP asymmetry is $A_{\mu e}$. A non zero value of $A_{\mu e}$ arises only in the presence of a non-trivial phase $\delta$, and its measurement plays a key role in distinguishing CP conserving from CP violating scenarios in the lepton sector.

In the standard three flavour framework, the appearance probability in matter can be approximated as
\begin{equation}
P(\nu_\mu \rightarrow \nu_e) \approx P_{\text{atm}} + P_{\text{sol}} + 2\sqrt{P_{\text{atm}} P_{\text{sol}}} \cos(\Delta_{32} + \delta),
\end{equation}
where $\Delta_{ij} = \dfrac{\Delta m^2_{ij} L}{4E}$ is the oscillation phase, with $L$ denoting the baseline and $E$ the neutrino energy. The atmospheric and solar contributions are given by
\begin{align}
\sqrt{P_{\text{atm}}} &= \sin\theta_{23} \sin 2\theta_{13} \frac{\sin[(\Delta_{31} - aL)]}{(\Delta_{31} - aL)} \Delta_{31}, \\
\sqrt{P_{\text{sol}}} &= \cos\theta_{23} \sin 2\theta_{12} \frac{\sin(aL)}{aL} \Delta_{21},
\end{align}
where $a = \dfrac{G_F N_e}{\sqrt{2}}$ denotes the matter potential, with $G_F$ and $N_e$ being the Fermi constant and the electron number density, respectively.

In the vacuum limit ($a \rightarrow 0$), the above expression simplifies, and the CP asymmetry takes the form
\begin{equation}
A_{\mu e} = \frac{2\sqrt{P_{\text{atm}} P_{\text{sol}}} \sin\Delta_{32} \sin\delta}
{P_{\text{atm}} + 2\sqrt{P_{\text{atm}} P_{\text{sol}}} \cos\Delta_{32} \cos\delta + P_{\text{sol}}}.
\end{equation}
The CP asymmetry parameter $A_{\mu e}$ depends explicitly on the Dirac CP phase $\delta$, the mass-squared differences $\Delta m_{21}^2$ and $\Delta m_{31}^2$, and the mixing angles $\theta_{12}$, $\theta_{13}$, and $\theta_{23}$. From our numerical analysis, we identify the allowed ranges of these parameters for which the model remains consistent with current oscillation and cosmological data. These model-predicted values are then used as inputs to compute the CP asymmetry parameter $A_{\mu e}$.
\begin{figure}[h]
    \centering
    % First row: three figures
    \begin{minipage}[b]{0.4\textwidth}
        \centering
        \includegraphics[width=\textwidth]{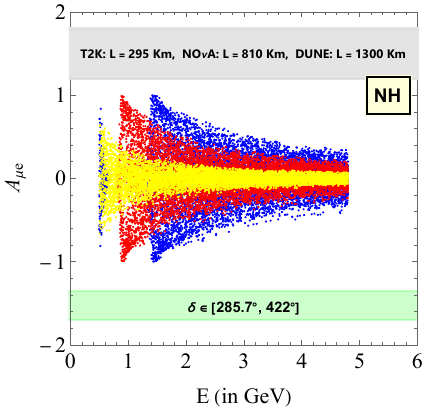}
    \end{minipage}
    \caption{The plot of $A_{\mu e}$ vs $E$. The `yellow', `red', and `blue' plot corresponds to T2K, NO$\nu$A and DUNE, respectively.}
\label{fig:cpasymmetry}
\end{figure}
In our model, the Dirac CP phase is constrained to lie within $\delta \in [285.7^\circ,\, 422^\circ]$, which effectively corresponds to $\delta \in [285.7^\circ,\, 360^\circ] \cup [0^\circ,\, 62^\circ]$ due to periodicity. The predicted ranges of $A_{\mu e}$ obtained by varying the neutrino energy in the range $E \in [0.5,\, 4.8]$~GeV, are highlighted below
\begin{align}
-0.69 < (A_{\mu e})_{\text{T2K}} &< 0.67 \quad (L = 295~\text{km}), \\
-0.99 < (A_{\mu e})_{\text{NO}\nu\text{A}} &< 0.98 \quad (L = 810~\text{km}), \\
-0.98 < (A_{\mu e})_{\text{DUNE}} &< 0.99 \quad (L = 1300~\text{km}).
\end{align}
Their current best fit values show a mild tension: T2K\,\cite{T2K:2011qtm} prefers $\delta \sim -0.69\pi$\,\cite{T2K:2025yoy} for NH, whereas NO$\nu$A\,\cite{NOvA:2007rmc} prefers $\delta \sim 0.82\pi$\,\cite{Catano-Mur:2022kyq} for NH. These values lie outside the range predicted by our model. However, due to current statistical limitations and uncertainties, the model remains consistent with existing data and can be tested in future high precision measurements. The upcoming DUNE\,\cite{DUNE:2021tad} experiment, with its longer baseline and better sensitivity, is expected to probe almost the entire predicted $A_{\mu e}$ range, providing a good test of the $\delta$ and other observables predicted by our model. To summarize the model predictions for $A_{\mu e}$, we present a plot of $A_{\mu e}$ as a function of $E$ in Fig.~\ref{fig:cpasymmetry}.
\section{Summary and Discussions}
\label{section 7}
We have presented a flavour driven neutrino mass model that stands out by generating a distinctive and predictive \emph{Echo 12–23 Texture}, realized from first principles within a Type-I+II seesaw framework. This elegant texture emerges naturally from an extended symmetry structure, $A_4 \times Z_3 \times Z_{10}$, which plays a crucial role in shaping the Yukawa sector and suppressing unwanted contributions via higher dimensional operators. Remarkably, the model accounts for the lepton mass hierarchy and predicts the entire neutrino mass matrix using just three real input parameters. The model makes sharp predictions for the three neutrino mass eigenvalues, as well as the Dirac and Majorana CP phases. Furthermore, it yields consistent predictions for the effective neutrino mass in light of $0\nu\beta\beta$ decay, the branching ratio of the dominant $\mu \rightarrow e\gamma$ decay, and the CP asymmetry parameter in the neutrino oscillations. This opens up prospects for verification in current and upcoming experiments. Overall, this framework offers a fresh path toward decoding the flavour puzzle in the lepton sector.

\section{Acknowledgement}
MD acknowledges financial support from the Council of Scientific and Industrial Research (CSIR), Government of India, through a NET–SRF grant No. 09/0059(15346)/2022-EMR-I, under which this work was carried out.

\section{Appendix}
\label{Appendix}
\begin{figure}
    \centering
    % First row: three figures
    \begin{minipage}[b]{0.35\textwidth}
        \centering
        \includegraphics[width=\textwidth]{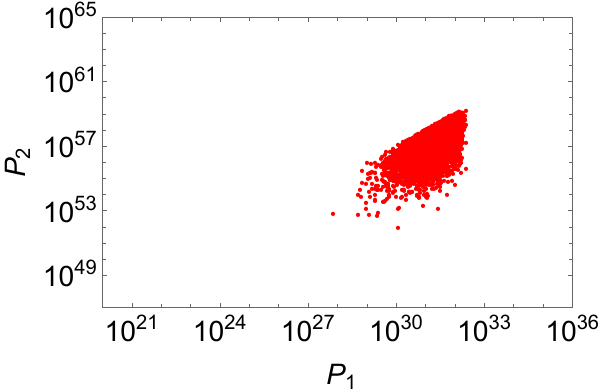}
        (a)
        \label{fig:1a}
    \end{minipage}
    \begin{minipage}[b]{0.35\textwidth}
        \centering
        \includegraphics[width=\textwidth]{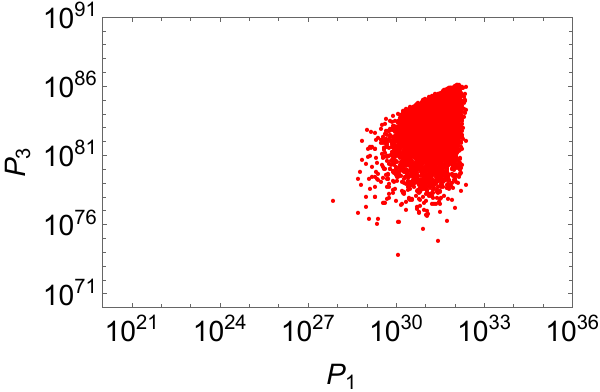}
        (b)
        \label{fig:1b}
    \end{minipage}
    \begin{minipage}[b]{0.36\textwidth}
        \centering
        \includegraphics[width=\textwidth]{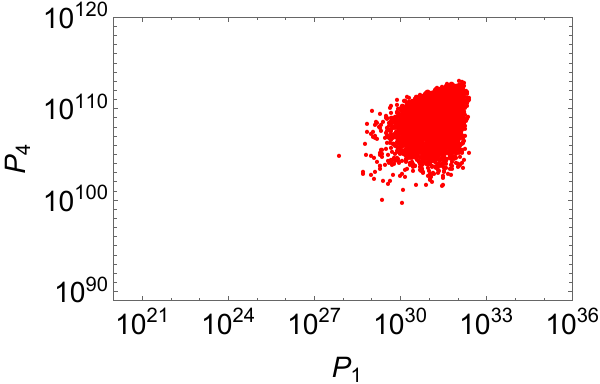}
       (c)
        \label{fig:1c}
    \end{minipage}
    \begin{minipage}[b]{0.36\textwidth}
        \centering
        \includegraphics[width=\textwidth]{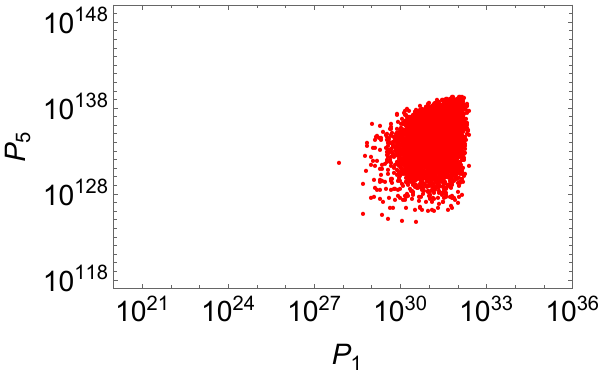}
        (d)
        \label{fig:1e}
    \end{minipage}
    \caption{The plots of the leading principal minors, where each $P_k$ has units of (eV)$^{2k}$.}
    \label{fig1}
\end{figure}

The scalar potential which is invariant under $\mathcal{G}$ is presented as shown
\begin{equation}
V= V(H)+ V(\Phi)+ V(\Delta)+ V(\chi)+ V(\kappa)+V(\sigma)+V(\rho)+ V(\zeta)+V(Interaction) + h.c.,\nonumber
\end{equation}
writing the terms explicitly
\begin{align}
V(H)&= -\mu^2_{H}(H^{\dagger}H)+ \lambda^{H}_1\,(H^{\dagger}H)(H^{\dagger}H)+\lambda^{H}_2\,(H^{\dagger}H)_{1'}(H^{\dagger}H)_{1''}+\lambda^{H}_3(H^{\dagger}H)_{3_s}(H^{\dagger}H)_{3_s}+\lambda^{H}_4\,(H^{\dagger}H)_{3_s}(H^{\dagger}\nonumber\\&\quad\,H )_{3_a} +\lambda^{H}_5(H^{\dagger}H)_{3_a}(H^{\dagger}H)_{3_a},\nonumber\\
V(\Phi)&= V(H \rightarrow \Phi; \lambda^{H}_i \rightarrow \lambda^{\Phi}_i),\nonumber\\
V(\Delta)&=-\mu^2_{\Delta}Tr(\Delta^{\dagger}\Delta)+ \lambda^{\Delta}_1 Tr(\Delta^{\dagger}\Delta)Tr(\Delta^{\dagger}\Delta)+\lambda^{\Delta}_2 Tr(\Delta^{\dagger}\Delta)_{1'} Tr(\Delta^{\dagger}\Delta)_{1''}+\lambda^{\Delta}_3 Tr(\Delta^{\dagger} \Delta)_{3_s} Tr(\Delta^{\dagger}\Delta)_{3_s}\, + \nonumber\\&\quad\,\lambda^{\Delta}_4 Tr(\Delta^{\dagger}\Delta)_{3_s} Tr(\Delta^{\dagger}\Delta)_{3_a}+\lambda^{\Delta}_5 Tr(\Delta^{\dagger}\Delta)_{3_a} Tr(\Delta^{\dagger}\Delta)_{3_a},\nonumber\\
V(\chi)&=  -\mu^2_{\chi}(\chi^{\dagger}\chi)+ \lambda^{\chi}(\chi^{\dagger}\chi)^2,\nonumber\\
V(\kappa)&=  -\mu^2_{\kappa}(\kappa^{\dagger}\kappa)+ \lambda^{\kappa}(\kappa^{\dagger}\kappa)^2,\nonumber\\
V(\sigma)&=  -\mu^2_{\sigma}(\sigma^{\dagger}\sigma)+ \lambda^{\sigma}(\sigma^{\dagger}\sigma)^2,\nonumber\\
V(\rho)&=  -\mu^2_{\rho}(\rho^{\dagger}\rho)+ \lambda^{\rho}(\rho^{\dagger}\rho)^2,\nonumber\\
V(\zeta)&=  -\mu^2_{\zeta}(\zeta^{\dagger}\zeta)+ \lambda^{\zeta}(\zeta^{\dagger}\zeta)^2,\nonumber\\
V(H,\Phi)&=\lambda^{H \Phi}_1(H^{\dagger}H)(\Phi^{\dagger}\Phi)+\lambda^{H \Phi}_2((H^{\dagger}H)_{1'}(\Phi^{\dagger}\Phi)_{1''}+(H^{\dagger}H)_{1''}(\Phi^{\dagger}\Phi)_{1'})+ \lambda^{H \Phi}_3(H^{\dagger}H)_{3_s}(\Phi^{\dagger}\Phi)_{3_s}+\lambda^{H \Phi}_4 ((H^{\dagger} \nonumber\\&\quad\,H)_{3_s}(\Phi^{\dagger}\Phi)_{3_a}+(H^{\dagger}H)_{3_a}(\Phi^{\dagger}\Phi)_{3_s})+\lambda^{H \Phi}_5(H^{\dagger}H)_{3_a}(\Phi^{\dagger}\Phi)_{3_a},\nonumber\\
V(H, \Delta)&= V(H, \Phi \rightarrow H,\Delta ; \lambda^{H \Phi}_i \rightarrow \lambda^{H \Delta}_i)+ \mu H^{T} i \sigma_2 \Delta^{T} H,\nonumber\\
V(\Phi,\Delta)&=V(H,\Phi \rightarrow \Phi,\Delta; \lambda^{H \Phi}_i \rightarrow \lambda^{\Phi \Delta}_i),\nonumber\\
V(H,(\chi, \kappa, \sigma, \rho, \zeta))&=(H^{\dagger}H)(\lambda^{H \chi} \chi^{\dagger}\chi + \lambda^{H \kappa} \kappa^{\dagger}\kappa + \lambda^{H \sigma} \sigma^{\dagger}\sigma + \lambda^{H \rho} \rho^{\dagger}\rho + \lambda^{H \zeta} \zeta^{\dagger}\zeta),\nonumber\\
V(\Phi,(\chi, \kappa, \sigma, \rho, \zeta))&= (\Phi^{\dagger}\Phi)(\lambda^{\Phi \chi} \chi^{\dagger}\chi + \lambda^{\Phi \kappa} \kappa^{\dagger}\kappa + \lambda^{\Phi \sigma} \sigma^{\dagger}\sigma + \lambda^{\Phi \rho} \rho^{\dagger}\rho + \lambda^{\Phi \zeta} \zeta^{\dagger}\zeta) ,\nonumber\\
V(\Delta,(\chi, \kappa, \sigma, \rho, \zeta))&= Tr(\Delta^{\dagger}\Delta)(\lambda^{\Delta \chi} \chi^{\dagger}\chi + \lambda^{\Delta \kappa} \kappa^{\dagger}\kappa + \lambda^{\Delta \sigma} \sigma^{\dagger}\sigma + \lambda^{\Delta \rho} \rho^{\dagger}\rho + \lambda^{\Delta \zeta} \zeta^{\dagger}\zeta).\nonumber
\end{align}
To verify whether the chosen vacuum alignment corresponds to a local minimum, two conditions must be satisfied. First, the first derivatives of the potential must vanish at the vacuum, i.e., the alignment must correspond to a stationary point. A local minimum is ensured by the positivity of the leading principal minors\,($P_i$) of the Hessian(excluding zeroes, associated with flat directions). In this model, there are 41 real scalar degrees of freedom. Hence, the Hessian is a $41 \times 41$ matrix. 
\\
First, we identify conditions under which all first derivatives vanish. The conditions are given below
\begin{align}
\lambda^H_ 1 & = -(8 v^2 _h\lambda_ 3^H + 
     v_ {\Delta} (9 v_ {\Delta}\lambda^{H\Delta} _ 1 +   4 v_ {\Delta}\lambda_ 3^{H\Delta} + 6\sqrt {2}\mu) +  9 (v^2 _ {\zeta}\lambda^{H\zeta} +  v^2 _ {\kappa}\lambda^{H\kappa} +   v^2 _ {\rho}\lambda^{H \rho} +  v^2 {\sigma}\lambda^{H\sigma} + v^2 _ {\chi}\lambda^{H\chi} -  2\mu^2 _H))/(18 v^2 _h), \nonumber \\
\lambda_1^{\Delta} & = -(v^3 _ {\Delta} (9\lambda^{\Delta} _ 2 +  4\lambda_ 3^{\Delta}) + 
     v^2 _ {\Phi} v_ {\Delta} (9\lambda_ 2^{\Phi\Delta} -  2\lambda_ 3^{\Phi\Delta}) +  v^2 _h (9 v_ {\Delta}\lambda_ 1^{H\Delta} + 
        4 v_ {\Delta}\lambda_ 3^{H\Delta} + 3\sqrt {2}\mu) +  9 v_ {\Delta}(v^2 _ {\zeta}\lambda^{\Delta\zeta} +  v^2 {\kappa}\lambda^{\Delta\kappa} +   v^2 _ {\rho}\lambda^{\Delta \rho}+ \nonumber\\&\quad\, v^2 _ {\sigma}\lambda^{\Delta \sigma} + v^2 _ {\chi}\lambda^{\Delta\chi} -  2\mu^2 _ {\Delta}))/(18 v^3 _ {\Delta}), \nonumber \\
\lambda^{H\Delta} _ 3 & = (4 v^2 _ {\Phi} (9\lambda_ 1^{H\Phi} - 2\lambda_ 3^{H\Phi}) - 
    9\sqrt {2} v_ {\Delta}\mu + (36 v^2 _ {\Phi} (v^2 _ {\Delta} \ (\lambda_ 1^{\Phi\Delta} - \lambda_ 2^{\Phi\Delta}) + 
           v^2 _ {\zeta}\lambda^{\Phi\zeta} + 
           v^2 _ {\kappa}\lambda^{\Phi\kappa} + 
           v^2 _ {\rho}\lambda^{\Phi\rho} + 
           v^2 _ {\sigma}\lambda^{\Phi\sigma} + 
           v^2 _ {\chi}\lambda^{\Phi\chi} - 
           2\mu^2 _ {\Phi})/\nonumber\\&\quad\,v^2 _h))/(18 v^2 _ {\Delta}), \nonumber \\
\lambda^{H\Phi} _ 1 & = (2\lambda^{H\Phi} _ 3/
     9) - (v^2 _ {\Delta} (\lambda_ 1^{\Phi\Delta} - \lambda_ 2^{\Phi\
\Delta}) + v^2 _ {\zeta}\lambda^{\Phi\zeta} + 
      v^2 _ {\kappa}\lambda^{\Phi\kappa} + 
      v^2 _ {\rho}\lambda^{\Phi \rho} + 
      v^2 _ {\sigma}\lambda^{\Phi\sigma} + 
      v^2 _ {\chi}\lambda^{\Phi\chi} - 2\mu^2 _ {\Phi})/
    v^2 _h, \nonumber \\
\lambda_ 2^{\Delta} & = (4 v^3 _ {\Delta}\lambda_ 3^{\Delta} +  v^2 _ {\Phi} v_ {\Delta} (9\lambda_ 2^{\Phi\Delta} -  2\lambda_ 3^{\Phi\Delta}) + 
    3 v^2 _h (2 v_ {\Delta}\lambda_ 3^{H\Delta} + \sqrt {2}\mu))/(9 v^3 _ {\Delta}), \nonumber \\
\lambda_ 3^{\Phi\Delta} & = 
 9 (4\lambda_ 2^{\Phi\Delta} + (v^2 _h (2 v_ {\Delta}\lambda_3^{H\Delta} + \sqrt {2}\mu)/(v^2 _ {\Phi} v_ {\Delta})))/8, \nonumber \\
\lambda^{\Phi} _ 2 & = (4\lambda^{\Phi} _ 3/
     9) + (v^2 _h v_ {\Delta} (2 v_ {\Delta}\lambda^{H\Delta} _ 3 + \sqrt {2}\mu))/(4 v^4 _ {\Phi}), \nonumber \\
\lambda^{H\chi} & = -(v^2 _ {\Delta}\lambda^{\Delta\chi} +  2 v^2 _ {\chi}\lambda^{\chi} - 2\mu^2 _ {\chi})/ v^2 _h, \nonumber \\
\lambda^{H\kappa} & = -(v^2 _ {\Delta}\lambda^{\Delta\kappa} +  2 v^2 _ {\kappa}\lambda^{\kappa} - 2\mu^2 _ {\kappa})/ v^2 _h, \nonumber \\
\lambda^{H\sigma} & = -(v^2 _ {\Delta}\lambda^{\Delta\sigma} + 2 v^2 _ {\sigma}\lambda^{\sigma} - 2\mu^2 _ {\sigma})/ v^2 _h, \nonumber \\
\lambda^{H \rho} & = -(v^2 _ {\Delta}\lambda^{\Delta \rho} +  2 v^2 _ {\rho}\lambda^{\rho} - 2\mu^2 _ {\rho})/ v^2 _h, \nonumber \\
\lambda^{H\zeta} & = -(v^2 _ {\Delta}\lambda^{\Delta\zeta} +  2 v^2 _ {\zeta}\lambda^{\zeta} - 2\mu^2 _ {\zeta})/ v^2 _h . \nonumber
\end{align}
Next, we evaluate the 41 dimensional Hessian matrix at the vacuum, which is given by
\[
\mathrm{Hess}_{ij} = \left.\frac{\partial^2 V}{\partial \phi_i \partial \phi_j}\right|_{0}.
\]
We substitute these 12 expressions into the Hessian and calculate its principal minors. We find that all the principal minors are non negative. Out of 41 leading principal minors, $P_6$ to $P_{41}$ correspond to zero, while the remaining five, as mentioned in the previous version of the manuscript, are listed below
\begin{align}
&P_1 = 4\mu^2_{H} 
- 2 \lambda_1^{H\Delta} v_\Delta^2 
- \frac{8}{9} \sqrt{2} \mu v_\Delta 
+ \frac{2 v_\Delta^2}{v_h^2}( 
    \lambda^{\Delta \zeta} v_\zeta^2 
  + \lambda^{\Delta \kappa} v_\kappa^2 
  + \lambda^{\Delta \rho} v_\rho^2 
  + \lambda^{\Delta \sigma} v_\sigma^2 
  + \lambda^{\Delta \chi} v_\chi^2 ) + \frac{4}{v_h^2}(\lambda_\zeta v_\zeta^4 
  - \mu_{\zeta}^2 v_\zeta^2 
  + \lambda^\kappa v_\kappa^4 
  - \mu^2_{\kappa} v_\kappa^2\,+ \nonumber\\&\quad\, 
  \lambda^\rho v_\rho^4 
  - \mu^2_{\rho} v_\rho^2 
  + \lambda^\sigma v_\sigma^4 
  - \mu^2_{\sigma} v_\sigma^2 
  + \lambda^\chi v_\chi^4 
  - \mu^2_{\chi} v_\chi^2 ), \\[5pt]
&P_2 = -\frac{2}{27 v_h^2} ( 
    9 v_\Delta^2 \lambda_2^{\Phi\Delta} + 2 v_h^2 \lambda_3^{H\Phi} ) [ v_h^2 ( 
        9 v_\Delta^2 \lambda_1^{H\Delta} + 4 \sqrt{2} v_\Delta \mu - 18 \mu^2_{H} 
    )  - 9 \,( 
        v_\Delta^2 \, ( 
            v_\zeta^2 \lambda^{\Delta\zeta} 
          + v_\kappa^2  \lambda^{\Delta\kappa}  
          + v_\rho^2 \lambda^{\Delta\rho} 
          + v_\sigma^2 \lambda^{\Delta\sigma} 
          + v_\chi^2 \lambda^{\Delta\chi} 
        )+ \nonumber\\&\quad\,  2 ( 
            v_\zeta^4 \lambda^\zeta 
          + v_\kappa^4 \lambda^\kappa 
          + v_\rho^4 \lambda^\rho 
          + v_\sigma^4 \lambda^\sigma 
          + v_\chi^4 \lambda^\chi 
          - v_\zeta^2 \mu^2_{\zeta} 
          - v_\kappa^2 \mu^2_{\kappa} 
          - v_\rho^2 \mu^2_{\rho} 
          - v_\sigma^2 \mu^2_{\sigma} 
          - v_\chi^2 \mu^2_{\chi} 
        )) ], \\[5pt]
&P_3 = ( 3 v_\Delta^2 \lambda_2^{\Phi\Delta} + 2 v_h^2 \lambda_3^{H\Phi})^2 [
    -2 v_\Delta^2 \lambda_1^{H\Delta}
    + \frac{2 v_\Delta^2}{v_h^2}( v_\zeta^2 \lambda^{\Delta\zeta} 
      + v_\kappa^2 \lambda^{\Delta\kappa} 
      + v_\rho^2 \lambda^{\Delta\rho} 
      + v_\sigma^2 \lambda^{\Delta\sigma} 
      + v_\chi^2 \lambda^{\Delta\chi})- \frac{8}{9} \sqrt{2} v_\Delta \mu + 4 \mu^2_{H} 
    + \frac{4}{v_h^2} ( v_\zeta^4 \lambda^\zeta \nonumber\\&\quad\, 
      + v_\kappa^4 \lambda^\kappa 
      + v_\rho^4 \lambda^\rho 
      + v_\sigma^4 \lambda^\sigma 
      + v_\chi^4 \lambda^\chi 
      - v_\zeta^2 \mu^2_{\zeta} 
      - v_\kappa^2 \mu^2_{\kappa}
      - v_\rho^2 \mu^2_{\rho} 
      - v_\sigma^2 \mu^2_{\sigma} 
      - v_\chi^2 \mu^2_{\chi} 
      )],\\[5pt]
& P_4 = (3 v_\Delta^2 \lambda_2^{\Phi\Delta} + 2 v_h^2 \lambda_3^{H\Phi})
[-9 v_\Delta^4 (\lambda_2^{\Phi\Delta})^2(
        -2 v_\Delta^2 \lambda_1^{H\Delta}
        + \frac{2 v_\Delta^2}{v_h^2} (
            v_\zeta^2 \lambda^{\Delta\zeta}
          + v_\kappa^2 \lambda^{\Delta\kappa}
          + v_\rho^2 \lambda^{\Delta\rho}
          + v_\sigma^2 \lambda^{\Delta\sigma}
          + v_\chi^2 \lambda^{\Delta\chi}
        ) - \frac{8}{9} \sqrt{2} v_\Delta \mu 
        + 4 \mu^2_{H} \nonumber\\&\quad\, + \frac{4}{v_h^2} (
            v_\zeta^4 \lambda^\zeta 
          + v_\kappa^4 \lambda^\kappa
          + v_\rho^4 \lambda^\rho
          + v_\sigma^4 \lambda^\sigma
          + v_\chi^4 \lambda^\chi
          - v_\zeta^2 \mu^2_{\zeta}
          - v_\kappa^2 \mu^2_{\kappa}
          - v_\rho^2 \mu^2_{\rho}
          - v_\sigma^2 \mu^2_{\sigma}
          - v_\chi^2 \mu^2_{\chi}
        )
    )
    + ((3 v_\Delta^2 \lambda_2^{\Phi\Delta} + 2 v_h^2 \lambda_3^{H\Phi})/3 )[
        -\frac{4 v_\Phi^2}{v_h^2} (
            v_\Delta^2 \nonumber\\&\quad\, (\lambda_1^{\Phi\Delta} - \lambda_2^{\Phi\Delta} )
          + v_\zeta^2 \lambda^{\Phi\zeta} 
          + v_\kappa^2 \lambda^{\Phi\kappa}
          + v_\rho^2 \lambda^{\Phi\rho}
          + v_\sigma^2 \lambda^{\Phi\sigma} 
          + v_\chi^2 \lambda^{\Phi\chi}
          - 2 \mu^2_{\Phi}
        )^2 
        + \,( 3 v_\Delta^2 \lambda_2^{\Phi\Delta} + \frac{4}{9} v_\Phi^2 \, (9 \lambda_1^{\Phi} + 4 \lambda_3^{\Phi}) )
        (
            -2 v_\Delta^2 \lambda_1^{H\Delta}
            + \frac{2 v_\Delta^2}{v_h^2} (\nonumber\\&\quad\, 
                v_\zeta^2  \lambda^{\Delta\zeta}
              + v_\kappa^2 \lambda^{\Delta\kappa}
              + v_\rho^2 \lambda^{\Delta\rho}
              + v_\sigma^2 \lambda^{\Delta\sigma}
              + v_\chi^2 \lambda^{\Delta\chi}
            )
            - \frac{8}{9} \sqrt{2} v_\Delta \mu
            + 4 \mu^2_{H} 
            + \frac{4}{v_h^2}(
                v_\zeta^4 \lambda^\zeta
              + v_\kappa^4 \lambda^\kappa
              + v_\rho^4 \lambda^\rho
              + v_\sigma^4 \lambda^\sigma
              + v_\chi^4 \lambda^\chi 
              -\, v_\zeta^2 \mu^2_{\zeta}
              - v_\kappa^2 \nonumber\\&\quad\,  \mu^2_{\kappa}
              - v_\rho^2 \mu^2_{\rho}
              - v_\sigma^2 \mu^2_{\sigma}
              - v_\chi^2 \mu^2_{\chi}
            ) ) ] ],\\[5pt]
&P_5 = \frac{2}{2187\, v_h^3}( 2 v_h^2 \lambda_3^{H\Phi} ( 9 v_\Delta^2 \lambda_2^{\Phi\Delta} + v_\Phi^2 (4 \lambda_3^{\Phi} - 3 \lambda_5^{\Phi}) ) + 9 v_\Phi^2 v_\Delta^2 \lambda_2^{\Phi\Delta}  (4 \lambda_3^{\Phi} - 3 \lambda_5^{\Phi}))( 243 v_h v_\Delta^4 (\lambda_2^{\Phi\Delta})^2 ( v_h^2( 9 v_\Delta^2 \lambda_1^{H\Delta} + 4 \sqrt{2} v_\Delta \mu \,- \nonumber\\[3pt]&\quad\,  18 \mu_H^2 ) - 9 ( v_\Delta^2 ( v_\zeta^2 \lambda^{\Delta\zeta} + v_\kappa^2 \lambda^{\Delta\kappa} + v_\rho^2 \lambda^{\Delta\rho} + v_\sigma^2 \lambda^{\Delta\sigma} + v_\chi^2 \lambda^{\Delta\chi}) + 2 (v_\zeta^4 \lambda^\zeta + v_\kappa^4 \lambda^\kappa + v_\rho^4 \lambda^\rho + v_\sigma^4 \lambda^\sigma + v_\chi^4 \lambda^\chi - v_\zeta^2 \mu_\zeta^2 - v_\kappa^2 \mu_\kappa^2 - v_\rho^2 \nonumber\\[4pt]&\quad\,  \mu_\rho^2 - v_\sigma^2 \mu_\sigma^2 - v_\chi^2 \mu_\chi^2 ) )) + ( 9 v_\Delta^2 \lambda_2^{\Phi\Delta} + 2 v_h^2 \lambda_3^{H\Phi} ) ( -162 v_h v_\Phi^2 ( v_\Delta^2 (\lambda_1^{\Phi\Delta} - \lambda_2^{\Phi\Delta}) + v_\zeta^2 \lambda^{\Phi\zeta} + v_\kappa^2 \lambda^{\Phi\kappa} + v_\rho^2 \lambda^{\Phi\rho} + v_\sigma^2 \lambda^{\Phi\sigma} + v_\chi^2 \lambda^{\Phi\chi} - \nonumber\\[4pt]&\quad\,  2 \mu_\Phi^2 )^2 + v_h ( 27 v_\Delta^2 \lambda_2^{\Phi\Delta} + 4 v_\Phi^2 (9 \lambda_1^{\Phi} + 4 \lambda_3^{\Phi}))\,\, ( 9 v_\Delta^2 (v_\zeta^2 \lambda^{\Delta\zeta} + v_\kappa^2 \lambda^{\Delta\kappa} + v_\rho^2 \lambda^{\Delta\rho} + v_\sigma^2 \lambda^{\Delta\sigma} + v_\chi^2 \lambda^{\Delta\chi} ) + v_h^2 ( -9 v_\Delta^2 \lambda_1^{H\Delta} - 4 \sqrt{2} v_\Delta \mu \nonumber\\[4pt]&\quad\, + 18 \mu_H^2 ) + 18 ( v_\zeta^4 \lambda^\zeta + v_\kappa^4 \lambda^\kappa + v_\rho^4 \lambda^\rho + v_\sigma^4 \lambda^\sigma + v_\chi^4 \lambda^\chi - v_\zeta^2 \mu_\zeta^2 - v_\kappa^2 \mu_\kappa^2 - v_\rho^2 \mu_\rho^2 - v_\sigma^2 \mu_\sigma^2 - v_\chi^2 \mu_\chi^2 ) ) ) ).
\end{align}
To ensure a true minimum of the potential, we require $P_{1-5} > 0$. From their expressions, we see that there are sufficient free parameters to guarantee this condition, which are 
$v_h, v_\phi, v_\Delta, v_\zeta, v_\kappa, v_\rho, v_\sigma, v_\chi,$ $\mu_1, \mu_H^2, \mu_\Delta^2, \mu_\zeta^2, \mu_\kappa^2, \mu_\rho^2, 
\mu_\sigma^2, \mu_\Phi^2, \mu_\chi^2,$ $\lambda_1^{H\Delta}, \lambda_1^\Phi, \lambda_1^{\Phi\Delta}, 
\lambda_2^{\Phi\Delta}, \lambda_3^{H\Phi}, \lambda_3^\Delta, 
\lambda_3^\Phi, \lambda_5^\Delta,$ $\lambda_5^\Phi, \lambda_5^{\Phi\Delta}, 
\lambda^{\Delta\zeta}, \lambda^{\Delta\kappa}, \lambda^{\Delta\rho}, 
\lambda^{\Delta\sigma}, \lambda^{\Delta\chi}, \lambda_\zeta, \lambda_\kappa, 
\lambda_\rho, \lambda_\sigma, \lambda_\chi, \lambda^{\Phi\zeta}, 
\lambda^{\Phi\kappa}, \lambda^{\Phi\rho}, \lambda^{\Phi\sigma}, \lambda^{\Phi\chi}$. 
\\
We perform a numerical scan by varying the vevs within the electroweak to TeV range. The dimensionless couplings $\lambda_i$ are varied uniformly in the range $[-1, 1]$, while the mass parameters $\mu^2$ are scanned within $[200, 10^{22}]~\text{eV}^2$. During this process, $\mu_1$ is varied in the range $[200, 10^8]~\text{eV}$, and the electroweak vev is fixed at $v_h = 246~\text{GeV}$. A sufficiently large set of random parameter points is generated to check $P_{1-5} > 0$. In Figure~\ref{fig1}, we show the plots of $P_{1-5}$, which confirm their positivity in our model.
It is important to mention that we have included all cross term contributions between the triplets $H, \Phi, \Delta$ and the singlet scalars both in the analytic conditions for vanishing first derivatives and in the numerical evaluation of the Hessian matrix, which captures the curvature of the scalar potential in all field directions. In our analysis, first five leading principal minors are positive, corresponding to the directions that determine the vacuum alignment, ensuring stability along these essential directions. The remaining directions correspond to flat directions, where the curvature vanishes and the potential remains constant. Therefore, the chosen vacuum alignment remains robust and corresponds to a minimum of the scalar potential.
\setlength{\bibsep}{1ex}
\biboptions{sort&compress}
\bibliography{references} % your cleaned .bib file

\begin{thebibliography}{10}
\expandafter\ifx\csname url\endcsname\relax
  \def\url#1{\texttt{#1}}\fi
\expandafter\ifx\csname urlprefix\endcsname\relax\def\urlprefix{URL }\fi
\expandafter\ifx\csname href\endcsname\relax
  \def\href#1#2{#2} \def\path#1{#1}\fi

\bibitem{Glashow:1961tr}
S.~L. Glashow, {Partial Symmetries of Weak Interactions}, Nucl. Phys. 22 (1961)
  579--588.
\newblock \href {https://doi.org/10.1016/0029-5582(61)90469-2}
  {\path{doi:10.1016/0029-5582(61)90469-2}}.

\bibitem{Weinberg:1967tq}
S.~Weinberg, {A Model of Leptons}, Phys. Rev. Lett. 19 (1967) 1264--1266.
\newblock \href {https://doi.org/10.1103/PhysRevLett.19.1264}
  {\path{doi:10.1103/PhysRevLett.19.1264}}.

\bibitem{Salam:1968rm}
A.~Salam, {Weak and Electromagnetic Interactions}, Conf. Proc. C 680519 (1968)
  367--377.
\newblock \href {https://doi.org/10.1142/9789812795915_0034}
  {\path{doi:10.1142/9789812795915_0034}}.

\bibitem{Gross:1973id}
D.~J. Gross, F.~Wilczek, {Ultraviolet Behavior of Nonabelian Gauge Theories},
  Phys. Rev. Lett. 30 (1973) 1343--1346.
\newblock \href {https://doi.org/10.1103/PhysRevLett.30.1343}
  {\path{doi:10.1103/PhysRevLett.30.1343}}.

\bibitem{Politzer:1973fx}
H.~D. Politzer, {Reliable Perturbative Results for Strong Interactions?}, Phys.
  Rev. Lett. 30 (1973) 1346--1349.
\newblock \href {https://doi.org/10.1103/PhysRevLett.30.1346}
  {\path{doi:10.1103/PhysRevLett.30.1346}}.

\bibitem{Cowan:1956rrn}
C.~L. Cowan, F.~Reines, F.~B. Harrison, H.~W. Kruse, A.~D. McGuire, {Detection
  of the free neutrino: A Confirmation}, Science 124 (1956) 103--104.
\newblock \href {https://doi.org/10.1126/science.124.3212.103}
  {\path{doi:10.1126/science.124.3212.103}}.

\bibitem{SNO:2002tuh}
Q.~R. Ahmad, et~al., {Direct evidence for neutrino flavor transformation from
  neutral current interactions in the Sudbury Neutrino Observatory}, Phys. Rev.
  Lett. 89 (2002) 011301.
\newblock \href {http://arxiv.org/abs/nucl-ex/0204008}
  {\path{arXiv:nucl-ex/0204008}}, \href
  {https://doi.org/10.1103/PhysRevLett.89.011301}
  {\path{doi:10.1103/PhysRevLett.89.011301}}.

\bibitem{KamLAND:2002uet}
K.~Eguchi, et~al., {First results from KamLAND: Evidence for reactor
  anti-neutrino disappearance}, Phys. Rev. Lett. 90 (2003) 021802.
\newblock \href {http://arxiv.org/abs/hep-ex/0212021}
  {\path{arXiv:hep-ex/0212021}}, \href
  {https://doi.org/10.1103/PhysRevLett.90.021802}
  {\path{doi:10.1103/PhysRevLett.90.021802}}.

\bibitem{Super-Kamiokande:1998uiq}
Y.~Fukuda, et~al., {Measurement of the flux and zenith angle distribution of
  upward through going muons by Super-Kamiokande}, Phys. Rev. Lett. 82 (1999)
  2644--2648.
\newblock \href {http://arxiv.org/abs/hep-ex/9812014}
  {\path{arXiv:hep-ex/9812014}}, \href
  {https://doi.org/10.1103/PhysRevLett.82.2644}
  {\path{doi:10.1103/PhysRevLett.82.2644}}.

\bibitem{Yoshimura:1978ex}
M.~Yoshimura, {Unified Gauge Theories and the Baryon Number of the Universe},
  Phys. Rev. Lett. 41 (1978) 281--284, [Erratum: Phys.Rev.Lett. 42, 746
  (1979)].
\newblock \href {https://doi.org/10.1103/PhysRevLett.41.281}
  {\path{doi:10.1103/PhysRevLett.41.281}}.

\bibitem{Akhmedov:1999tm}
E.~K. Akhmedov, G.~C. Branco, M.~N. Rebelo, {Seesaw mechanism and structure of
  neutrino mass matrix}, Phys. Lett. B 478 (2000) 215--223.
\newblock \href {http://arxiv.org/abs/hep-ph/9911364}
  {\path{arXiv:hep-ph/9911364}}, \href
  {https://doi.org/10.1016/S0370-2693(00)00282-3}
  {\path{doi:10.1016/S0370-2693(00)00282-3}}.

\bibitem{Mohapatra:2004zh}
R.~N. Mohapatra, {Seesaw mechanism and its implications}, in: {SEESAW25:
  International Conference on the Seesaw Mechanism and the Neutrino Mass},
  2004, pp. 29--44.
\newblock \href {http://arxiv.org/abs/hep-ph/0412379}
  {\path{arXiv:hep-ph/0412379}}, \href
  {https://doi.org/10.1142/9789812702210_0003}
  {\path{doi:10.1142/9789812702210_0003}}.

\bibitem{King:2003jb}
S.~F. King, {Neutrino mass models}, Rept. Prog. Phys. 67 (2004) 107--158.
\newblock \href {http://arxiv.org/abs/hep-ph/0310204}
  {\path{arXiv:hep-ph/0310204}}, \href
  {https://doi.org/10.1088/0034-4885/67/2/R01}
  {\path{doi:10.1088/0034-4885/67/2/R01}}.

\bibitem{Cheng:1980qt}
T.~P. Cheng, L.-F. Li, {Neutrino Masses, Mixings and Oscillations in SU(2) x
  U(1) Models of Electroweak Interactions}, Phys. Rev. D 22 (1980) 2860.
\newblock \href {https://doi.org/10.1103/PhysRevD.22.2860}
  {\path{doi:10.1103/PhysRevD.22.2860}}.

\bibitem{Cai:2017mow}
Y.~Cai, T.~Han, T.~Li, R.~Ruiz, {Lepton Number Violation: Seesaw Models and
  Their Collider Tests}, Front. in Phys. 6 (2018) 40.
\newblock \href {http://arxiv.org/abs/1711.02180} {\path{arXiv:1711.02180}},
  \href {https://doi.org/10.3389/fphy.2018.00040}
  {\path{doi:10.3389/fphy.2018.00040}}.

\bibitem{Akhmedov:2006de}
E.~K. Akhmedov, M.~Frigerio, {Interplay of type I and type II seesaw
  contributions to neutrino mass}, JHEP 01 (2007) 043.
\newblock \href {http://arxiv.org/abs/hep-ph/0609046}
  {\path{arXiv:hep-ph/0609046}}, \href
  {https://doi.org/10.1088/1126-6708/2007/01/043}
  {\path{doi:10.1088/1126-6708/2007/01/043}}.

\bibitem{Verma:2018lro}
S.~Verma, M.~Kashav, S.~Bhardwaj, {Highly predictive and testable $A_4$ flavor
  model within type-I and II seesaw framework and associated phenomenology},
  Nucl. Phys. B 946 (2019) 114704.
\newblock \href {http://arxiv.org/abs/1811.06249} {\path{arXiv:1811.06249}},
  \href {https://doi.org/10.1016/j.nuclphysb.2019.114704}
  {\path{doi:10.1016/j.nuclphysb.2019.114704}}.

\bibitem{Singh:2022nmk}
L.~Singh, Tapender, M.~Kashav, S.~Verma, {Trimaximal mixing and extended magic
  symmetry in a model of neutrino mass matrix}, EPL 142~(6) (2023) 64002.
\newblock \href {http://arxiv.org/abs/2207.13328} {\path{arXiv:2207.13328}},
  \href {https://doi.org/10.1209/0295-5075/acdb97}
  {\path{doi:10.1209/0295-5075/acdb97}}.

\bibitem{Ramond:1993kv}
P.~Ramond, R.~G. Roberts, G.~G. Ross, {Stitching the Yukawa quilt}, Nucl. Phys.
  B 406 (1993) 19--42.
\newblock \href {http://arxiv.org/abs/hep-ph/9303320}
  {\path{arXiv:hep-ph/9303320}}, \href
  {https://doi.org/10.1016/0550-3213(93)90159-M}
  {\path{doi:10.1016/0550-3213(93)90159-M}}.

\bibitem{Fritzsch:2011qv}
H.~Fritzsch, Z.-z. Xing, S.~Zhou, {Two-zero Textures of the Majorana Neutrino
  Mass Matrix and Current Experimental Tests}, JHEP 09 (2011) 083.
\newblock \href {http://arxiv.org/abs/1108.4534} {\path{arXiv:1108.4534}},
  \href {https://doi.org/10.1007/JHEP09(2011)083}
  {\path{doi:10.1007/JHEP09(2011)083}}.

\bibitem{Ludl:2014axa}
P.~O. Ludl, W.~Grimus, {A complete survey of texture zeros in the lepton mass
  matrices}, JHEP 07 (2014) 090, [Erratum: JHEP 10, 126 (2014)].
\newblock \href {http://arxiv.org/abs/1406.3546} {\path{arXiv:1406.3546}},
  \href {https://doi.org/10.1007/JHEP07(2014)090}
  {\path{doi:10.1007/JHEP07(2014)090}}.

\bibitem{Harrison:2002er}
P.~F. Harrison, D.~H. Perkins, W.~G. Scott, {Tri-bimaximal mixing and the
  neutrino oscillation data}, Phys. Lett. B 530 (2002) 167.
\newblock \href {http://arxiv.org/abs/hep-ph/0202074}
  {\path{arXiv:hep-ph/0202074}}, \href
  {https://doi.org/10.1016/S0370-2693(02)01336-9}
  {\path{doi:10.1016/S0370-2693(02)01336-9}}.

\bibitem{Xing:2022uax}
Z.-z. Xing, {The \textmu{}\textendash{}\ensuremath{\tau} reflection symmetry of
  Majorana neutrinos $^{*}$}, Rept. Prog. Phys. 86~(7) (2023) 076201.
\newblock \href {http://arxiv.org/abs/2210.11922} {\path{arXiv:2210.11922}},
  \href {https://doi.org/10.1088/1361-6633/acd8ce}
  {\path{doi:10.1088/1361-6633/acd8ce}}.

\bibitem{Ma:2001dn}
E.~Ma, G.~Rajasekaran, {Softly broken A(4) symmetry for nearly degenerate
  neutrino masses}, Phys. Rev. D 64 (2001) 113012.
\newblock \href {http://arxiv.org/abs/hep-ph/0106291}
  {\path{arXiv:hep-ph/0106291}}, \href
  {https://doi.org/10.1103/PhysRevD.64.113012}
  {\path{doi:10.1103/PhysRevD.64.113012}}.

\bibitem{King:2006np}
S.~F. King, M.~Malinsky, {A(4) family symmetry and quark-lepton unification},
  Phys. Lett. B 645 (2007) 351--357.
\newblock \href {http://arxiv.org/abs/hep-ph/0610250}
  {\path{arXiv:hep-ph/0610250}}, \href
  {https://doi.org/10.1016/j.physletb.2006.12.006}
  {\path{doi:10.1016/j.physletb.2006.12.006}}.

\bibitem{Altarelli:2010gt}
G.~Altarelli, F.~Feruglio, {Discrete Flavor Symmetries and Models of Neutrino
  Mixing}, Rev. Mod. Phys. 82 (2010) 2701--2729.
\newblock \href {http://arxiv.org/abs/1002.0211} {\path{arXiv:1002.0211}},
  \href {https://doi.org/10.1103/RevModPhys.82.2701}
  {\path{doi:10.1103/RevModPhys.82.2701}}.

\bibitem{King:2013eh}
S.~F. King, C.~Luhn, {Neutrino Mass and Mixing with Discrete Symmetry}, Rept.
  Prog. Phys. 76 (2013) 056201.
\newblock \href {http://arxiv.org/abs/1301.1340} {\path{arXiv:1301.1340}},
  \href {https://doi.org/10.1088/0034-4885/76/5/056201}
  {\path{doi:10.1088/0034-4885/76/5/056201}}.

\bibitem{Ma:2004yx}
E.~Ma, {Lepton family symmetry and neutrino mass matrix}, Mod. Phys. Lett. A 19
  (2004) 577--582.
\newblock \href {http://arxiv.org/abs/hep-ph/0401025}
  {\path{arXiv:hep-ph/0401025}}, \href
  {https://doi.org/10.1142/S0217732304013374}
  {\path{doi:10.1142/S0217732304013374}}.

\bibitem{Hu:2006wk}
B.~Hu, F.~Wu, Y.-L. Wu, {Z(3) Symmetry and Neutrino Mixing in Type II Seesaw},
  Phys. Rev. D 75 (2007) 113003.
\newblock \href {http://arxiv.org/abs/hep-ph/0612344}
  {\path{arXiv:hep-ph/0612344}}, \href
  {https://doi.org/10.1103/PhysRevD.75.113003}
  {\path{doi:10.1103/PhysRevD.75.113003}}.

\bibitem{CarcamoHernandez:2020udg}
A.~E. C\'arcamo~Hern\'andez, I.~de~Medeiros~Varzielas, {$\Delta(27)$ framework
  for cobimaximal neutrino mixing models}, Phys. Lett. B 806 (2020) 135491.
\newblock \href {http://arxiv.org/abs/2003.01134} {\path{arXiv:2003.01134}},
  \href {https://doi.org/10.1016/j.physletb.2020.135491}
  {\path{doi:10.1016/j.physletb.2020.135491}}.

\bibitem{Dey:2023rht}
M.~Dey, S.~Roy, {A realistic neutrino mixing scheme arising from A$_{4}$
  symmetry}, EPL 147 (2024) 14002.
\newblock \href {http://arxiv.org/abs/2304.07259} {\path{arXiv:2304.07259}},
  \href {https://doi.org/10.1209/0295-5075/ad57eb}
  {\path{doi:10.1209/0295-5075/ad57eb}}.

\bibitem{Dey:2024ctx}
M.~Dey, S.~Roy, {Revisiting the Dirac nature of neutrinos in the light of
  \ensuremath{\Delta}(27) and cyclic symmetries}, J. Phys. G 52~(2) (2025)
  025005.
\newblock \href {http://arxiv.org/abs/2403.12461} {\path{arXiv:2403.12461}},
  \href {https://doi.org/10.1088/1361-6471/ad9ec9}
  {\path{doi:10.1088/1361-6471/ad9ec9}}.

\bibitem{Van:2024qrs}
V.~V. Van, {$A_4 \times Z_2 \times Z_4$ flavor symmetry model for neutrino
  oscillation phenomenology}, Rev. Mex. Fis. 70~(6) (2024) 060801.
\newblock \href {https://doi.org/10.31349/RevMexFis.70.060801}
  {\path{doi:10.31349/RevMexFis.70.060801}}.

\bibitem{Altarelli:2005yx}
G.~Altarelli, F.~Feruglio, {Tri-bimaximal neutrino mixing, A(4) and the modular
  symmetry}, Nucl. Phys. B 741 (2006) 215--235.
\newblock \href {http://arxiv.org/abs/hep-ph/0512103}
  {\path{arXiv:hep-ph/0512103}}, \href
  {https://doi.org/10.1016/j.nuclphysb.2006.02.015}
  {\path{doi:10.1016/j.nuclphysb.2006.02.015}}.

\bibitem{ParticleDataGroup:2018ovx}
M.~Tanabashi, et~al., {Review of Particle Physics}, Phys. Rev. D 98~(3) (2018)
  030001.
\newblock \href {https://doi.org/10.1103/PhysRevD.98.030001}
  {\path{doi:10.1103/PhysRevD.98.030001}}.

\bibitem{ParticleDataGroup:2022pth}
R.~L. Workman, et~al., {Review of Particle Physics}, PTEP 2022 (2022) 083C01.
\newblock \href {https://doi.org/10.1093/ptep/ptac097}
  {\path{doi:10.1093/ptep/ptac097}}.

\bibitem{Maki:1962mu}
Z.~Maki, M.~Nakagawa, S.~Sakata, {Remarks on the unified model of elementary
  particles}, Prog. Theor. Phys. 28 (1962) 870--880.
\newblock \href {https://doi.org/10.1143/PTP.28.870}
  {\path{doi:10.1143/PTP.28.870}}.

\bibitem{Esteban:2024eli}
I.~Esteban, M.~C. Gonzalez-Garcia, M.~Maltoni, I.~Martinez-Soler, J.~a.~P.
  Pinheiro, T.~Schwetz, {NuFit-6.0: updated global analysis of three-flavor
  neutrino oscillations}, JHEP 12 (2024) 216.
\newblock \href {http://arxiv.org/abs/2410.05380} {\path{arXiv:2410.05380}},
  \href {https://doi.org/10.1007/JHEP12(2024)216}
  {\path{doi:10.1007/JHEP12(2024)216}}.

\bibitem{Planck:2018vyg}
N.~Aghanim, et~al., {Planck 2018 results. VI. Cosmological parameters}, Astron.
  Astrophys. 641 (2020) A6, [Erratum: Astron.Astrophys. 652, C4 (2021)].
\newblock \href {http://arxiv.org/abs/1807.06209} {\path{arXiv:1807.06209}},
  \href {https://doi.org/10.1051/0004-6361/201833910}
  {\path{doi:10.1051/0004-6361/201833910}}.

\bibitem{Schechter:1981bd}
J.~Schechter, J.~W.~F. Valle, {Neutrinoless Double beta Decay in SU(2) x U(1)
  Theories}, Phys. Rev. D 25 (1982) 2951.
\newblock \href {https://doi.org/10.1103/PhysRevD.25.2951}
  {\path{doi:10.1103/PhysRevD.25.2951}}.

\bibitem{Blennow:2010th}
M.~Blennow, E.~Fernandez-Martinez, J.~Lopez-Pavon, J.~Menendez, {Neutrinoless
  double beta decay in seesaw models}, JHEP 07 (2010) 096.
\newblock \href {http://arxiv.org/abs/1005.3240} {\path{arXiv:1005.3240}},
  \href {https://doi.org/10.1007/JHEP07(2010)096}
  {\path{doi:10.1007/JHEP07(2010)096}}.

\bibitem{Barabash:2023dwc}
A.~Barabash, {Double Beta Decay Experiments: Recent Achievements and Future
  Prospects}, Universe 9~(6) (2023) 290.
\newblock \href {https://doi.org/10.3390/universe9060290}
  {\path{doi:10.3390/universe9060290}}.

\bibitem{CUORE:2021mvw}
D.~Q. Adams, et~al., {Search for Majorana neutrinos exploiting millikelvin
  cryogenics with CUORE}, Nature 604~(7904) (2022) 53--58.
\newblock \href {http://arxiv.org/abs/2104.06906} {\path{arXiv:2104.06906}},
  \href {https://doi.org/10.1038/s41586-022-04497-4}
  {\path{doi:10.1038/s41586-022-04497-4}}.

\bibitem{EXO-200:2019rkq}
G.~Anton, et~al., {Search for Neutrinoless Double-$\beta$ Decay with the
  Complete EXO-200 Dataset}, Phys. Rev. Lett. 123~(16) (2019) 161802.
\newblock \href {http://arxiv.org/abs/1906.02723} {\path{arXiv:1906.02723}},
  \href {https://doi.org/10.1103/PhysRevLett.123.161802}
  {\path{doi:10.1103/PhysRevLett.123.161802}}.

\bibitem{GERDA:2020xhi}
M.~Agostini, et~al., {Final Results of GERDA on the Search for Neutrinoless
  Double-$\beta$ Decay}, Phys. Rev. Lett. 125~(25) (2020) 252502.
\newblock \href {http://arxiv.org/abs/2009.06079} {\path{arXiv:2009.06079}},
  \href {https://doi.org/10.1103/PhysRevLett.125.252502}
  {\path{doi:10.1103/PhysRevLett.125.252502}}.

\bibitem{KamLAND-Zen:2016pfg}
A.~Gando, et~al., {Search for Majorana Neutrinos near the Inverted Mass
  Hierarchy Region with KamLAND-Zen}, Phys. Rev. Lett. 117~(8) (2016) 082503,
  [Addendum: Phys.Rev.Lett. 117, 109903 (2016)].
\newblock \href {http://arxiv.org/abs/1605.02889} {\path{arXiv:1605.02889}},
  \href {https://doi.org/10.1103/PhysRevLett.117.082503}
  {\path{doi:10.1103/PhysRevLett.117.082503}}.

\bibitem{KamLAND-Zen:2022tow}
S.~Abe, et~al., {Search for the Majorana Nature of Neutrinos in the Inverted
  Mass Ordering Region with KamLAND-Zen}, Phys. Rev. Lett. 130~(5) (2023)
  051801.
\newblock \href {http://arxiv.org/abs/2203.02139} {\path{arXiv:2203.02139}},
  \href {https://doi.org/10.1103/PhysRevLett.130.051801}
  {\path{doi:10.1103/PhysRevLett.130.051801}}.

\bibitem{LEGEND:2021bnm}
N.~Abgrall, et~al., {The Large Enriched Germanium Experiment for Neutrinoless
  $\beta\beta$ Decay}: {LEGEND-1000 Preconceptual Design Report} (7 2021).
\newblock \href {http://arxiv.org/abs/2107.11462} {\path{arXiv:2107.11462}}.

\bibitem{Calibbi:2017uvl}
L.~Calibbi, G.~Signorelli, {Charged Lepton Flavour Violation: An Experimental
  and Theoretical Introduction}, Riv. Nuovo Cim. 41~(2) (2018) 71--174.
\newblock \href {http://arxiv.org/abs/1709.00294} {\path{arXiv:1709.00294}},
  \href {https://doi.org/10.1393/ncr/i2018-10144-0}
  {\path{doi:10.1393/ncr/i2018-10144-0}}.

\bibitem{Ardu:2022sbt}
M.~Ardu, G.~Pezzullo, {Introduction to Charged Lepton Flavor Violation},
  Universe 8~(6) (2022) 299.
\newblock \href {http://arxiv.org/abs/2204.08220} {\path{arXiv:2204.08220}},
  \href {https://doi.org/10.3390/universe8060299}
  {\path{doi:10.3390/universe8060299}}.

\bibitem{Cei:2014jtm}
F.~Cei, D.~Nicolo, {Lepton Flavour Violation Experiments}, Adv. High Energy
  Phys. 2014 (2014) 282915.
\newblock \href {https://doi.org/10.1155/2014/282915}
  {\path{doi:10.1155/2014/282915}}.

\bibitem{Matsuzaki:2008ik}
A.~Matsuzaki, H.~Tanaka, {Lepton Flavor Violating tau ---\ensuremath{>} 3mu
  Decay in Type-III Two Higgs Doublet Model}, Phys. Rev. D 79 (2009) 015006.
\newblock \href {http://arxiv.org/abs/0809.3072} {\path{arXiv:0809.3072}},
  \href {https://doi.org/10.1103/PhysRevD.79.015006}
  {\path{doi:10.1103/PhysRevD.79.015006}}.

\bibitem{Akeroyd:2009nu}
A.~G. Akeroyd, M.~Aoki, H.~Sugiyama, {Lepton Flavour Violating Decays tau
  ---\ensuremath{>} anti-l ll and mu ---\ensuremath{>} e gamma in the Higgs
  Triplet Model}, Phys. Rev. D 79 (2009) 113010.
\newblock \href {http://arxiv.org/abs/0904.3640} {\path{arXiv:0904.3640}},
  \href {https://doi.org/10.1103/PhysRevD.79.113010}
  {\path{doi:10.1103/PhysRevD.79.113010}}.

\bibitem{PhysRevD.89.013008}
A.~Celis, V.~Cirigliano, E.~Passemar,
  \href{https://link.aps.org/doi/10.1103/PhysRevD.89.013008}{Lepton flavor
  violation in the higgs sector and the role of hadronic
  $\ensuremath{\tau}$-lepton decays}, Phys. Rev. D 89 (2014) 013008.
\newblock \href {https://doi.org/10.1103/PhysRevD.89.013008}
  {\path{doi:10.1103/PhysRevD.89.013008}}.
\newline\urlprefix\url{https://link.aps.org/doi/10.1103/PhysRevD.89.013008}

\bibitem{Hue:2013uw}
L.~T. Hue, D.~T. Huong, H.~N. Long, {Lepton flavor violating processes $\tau
  \to \mu\gamma$, $\tau \to 3\mu$ and $Z \to \mu\tau$ in the Supersymmetric
  economical 3-3-1 model}, Nucl. Phys. B 873 (2013) 207--247.
\newblock \href {http://arxiv.org/abs/1301.4652} {\path{arXiv:1301.4652}},
  \href {https://doi.org/10.1016/j.nuclphysb.2013.04.014}
  {\path{doi:10.1016/j.nuclphysb.2013.04.014}}.

\bibitem{Dinh:2013vya}
D.~N. Dinh, S.~T. Petcov, {Lepton Flavor Violating $\tau$ Decays in TeV Scale
  Type I See-Saw and Higgs Triplet Models}, JHEP 09 (2013) 086.
\newblock \href {http://arxiv.org/abs/1308.4311} {\path{arXiv:1308.4311}},
  \href {https://doi.org/10.1007/JHEP09(2013)086}
  {\path{doi:10.1007/JHEP09(2013)086}}.

\bibitem{Omura:2015xcg}
Y.~Omura, E.~Senaha, K.~Tobe, {$\tau$- and $\mu$-physics in a general two Higgs
  doublet model with $\mu-\tau$ flavor violation}, Phys. Rev. D 94~(5) (2016)
  055019.
\newblock \href {http://arxiv.org/abs/1511.08880} {\path{arXiv:1511.08880}},
  \href {https://doi.org/10.1103/PhysRevD.94.055019}
  {\path{doi:10.1103/PhysRevD.94.055019}}.

\bibitem{Zhou:2016ynv}
H.~Zhou, R.-Y. Zhang, L.~Han, W.-G. Ma, L.~Guo, C.~Chen, {Searching for $\tau
  \rightarrow \mu \gamma$ lepton-flavor-violating decay at super charm-tau
  factory}, Eur. Phys. J. C 76~(8) (2016) 421.
\newblock \href {http://arxiv.org/abs/1602.01181} {\path{arXiv:1602.01181}},
  \href {https://doi.org/10.1140/epjc/s10052-016-4251-1}
  {\path{doi:10.1140/epjc/s10052-016-4251-1}}.

\bibitem{PhysRevD.99.035020}
C.~O. Dib, T.~Gutsche, S.~G. Kovalenko, V.~E. Lyubovitskij, I.~Schmidt,
  \href{https://link.aps.org/doi/10.1103/PhysRevD.99.035020}{Bounds on lepton
  flavor violating physics and decays of neutral mesons from
  $\ensuremath{\tau}(\ensuremath{\mu})\ensuremath{\rightarrow}3\ensuremath{\ell}$,
  $\ensuremath{\ell}\ensuremath{\gamma}\ensuremath{\gamma}$-decays}, Phys. Rev.
  D 99 (2019) 035020.
\newblock \href {https://doi.org/10.1103/PhysRevD.99.035020}
  {\path{doi:10.1103/PhysRevD.99.035020}}.
\newline\urlprefix\url{https://link.aps.org/doi/10.1103/PhysRevD.99.035020}

\bibitem{Hernandez-Tome:2018fbq}
G.~Hern\'andez-Tom\'e, G.~L\'opez~Castro, P.~Roig, {Flavor violating leptonic
  decays of $\tau$ and $\mu$ leptons in the Standard Model with massive
  neutrinos}, Eur. Phys. J. C 79~(1) (2019) 84, [Erratum: Eur.Phys.J.C 80, 438
  (2020)].
\newblock \href {http://arxiv.org/abs/1807.06050} {\path{arXiv:1807.06050}},
  \href {https://doi.org/10.1140/epjc/s10052-019-6563-4}
  {\path{doi:10.1140/epjc/s10052-019-6563-4}}.

\bibitem{Calcuttawala:2018wgo}
Z.~Calcuttawala, A.~Kundu, S.~Nandi, S.~Kumar~Patra, {New physics with the
  lepton flavor violating decay $\tau\to 3\mu$}, Phys. Rev. D 97~(9) (2018)
  095009.
\newblock \href {http://arxiv.org/abs/1802.09218} {\path{arXiv:1802.09218}},
  \href {https://doi.org/10.1103/PhysRevD.97.095009}
  {\path{doi:10.1103/PhysRevD.97.095009}}.

\bibitem{Ferreira:2019qpf}
M.~M. Ferreira, T.~B. de~Melo, S.~Kovalenko, P.~R.~D. Pinheiro, F.~S. Queiroz,
  {Lepton Flavor Violation and Collider Searches in a Type I + II Seesaw
  Model}, Eur. Phys. J. C 79~(11) (2019) 955.
\newblock \href {http://arxiv.org/abs/1903.07634} {\path{arXiv:1903.07634}},
  \href {https://doi.org/10.1140/epjc/s10052-019-7422-z}
  {\path{doi:10.1140/epjc/s10052-019-7422-z}}.

\bibitem{MEG:2016leq}
A.~M. Baldini, et~al., {Search for the lepton flavour violating decay $\mu ^+
  \rightarrow e + \gamma $ with the full dataset of the MEG experiment}, Eur.
  Phys. J. C 76~(8) (2016) 434.
\newblock \href {http://arxiv.org/abs/1605.05081} {\path{arXiv:1605.05081}},
  \href {https://doi.org/10.1140/epjc/s10052-016-4271-x}
  {\path{doi:10.1140/epjc/s10052-016-4271-x}}.

\bibitem{Meucci:2022qbh}
M.~Meucci, {MEG II experiment status and prospect}, PoS NuFact2021 (2022) 120.
\newblock \href {http://arxiv.org/abs/2201.08200} {\path{arXiv:2201.08200}},
  \href {https://doi.org/10.22323/1.402.0120} {\path{doi:10.22323/1.402.0120}}.

\bibitem{MEGII:2023ltw}
K.~Afanaciev, et~al., {A search for $\mu ^+ \rightarrow e + \gamma $ with the
  first dataset of the MEG~II experiment}, Eur. Phys. J. C 84~(3) (2024) 216.
\newblock \href {http://arxiv.org/abs/2310.12614} {\path{arXiv:2310.12614}},
  \href {https://doi.org/10.1140/epjc/s10052-024-12416-2}
  {\path{doi:10.1140/epjc/s10052-024-12416-2}}.

\bibitem{Francesconi:2024bpo}
M.~Francesconi, {Status of the MEG II experiment}, PoS EPS-HEP2023 (2024) 355.
\newblock \href {https://doi.org/10.22323/1.449.0355}
  {\path{doi:10.22323/1.449.0355}}.

\bibitem{Bilenky:1977du}
S.~M. Bilenky, S.~T. Petcov, B.~Pontecorvo, {Lepton Mixing, mu --\ensuremath{>}
  e + gamma Decay and Neutrino Oscillations}, Phys. Lett. B 67 (1977) 309.
\newblock \href {https://doi.org/10.1016/0370-2693(77)90379-3}
  {\path{doi:10.1016/0370-2693(77)90379-3}}.

\bibitem{Ilakovac:1994kj}
A.~Ilakovac, A.~Pilaftsis, {Flavor violating charged lepton decays in
  seesaw-type models}, Nucl. Phys. B 437 (1995) 491.
\newblock \href {http://arxiv.org/abs/hep-ph/9403398}
  {\path{arXiv:hep-ph/9403398}}, \href
  {https://doi.org/10.1016/0550-3213(94)00567-X}
  {\path{doi:10.1016/0550-3213(94)00567-X}}.

\bibitem{Tommasini:1995ii}
D.~Tommasini, G.~Barenboim, J.~Bernabeu, C.~Jarlskog, {Nondecoupling of heavy
  neutrinos and lepton flavor violation}, Nucl. Phys. B 444 (1995) 451--467.
\newblock \href {http://arxiv.org/abs/hep-ph/9503228}
  {\path{arXiv:hep-ph/9503228}}, \href
  {https://doi.org/10.1016/0550-3213(95)00201-3}
  {\path{doi:10.1016/0550-3213(95)00201-3}}.

\bibitem{Forero:2011pc}
D.~V. Forero, S.~Morisi, M.~Tortola, J.~W.~F. Valle, {Lepton flavor violation
  and non-unitary lepton mixing in low-scale type-I seesaw}, JHEP 09 (2011)
  142.
\newblock \href {http://arxiv.org/abs/1107.6009} {\path{arXiv:1107.6009}},
  \href {https://doi.org/10.1007/JHEP09(2011)142}
  {\path{doi:10.1007/JHEP09(2011)142}}.

\bibitem{Datta:2021zzf}
A.~Datta, B.~Karmakar, A.~Sil, {Flavored leptogenesis and neutrino mass with
  A$_{4}$ symmetry}, JHEP 12 (2021) 051.
\newblock \href {http://arxiv.org/abs/2106.06773} {\path{arXiv:2106.06773}},
  \href {https://doi.org/10.1007/JHEP12(2021)051}
  {\path{doi:10.1007/JHEP12(2021)051}}.

\bibitem{Dey:2023bfa}
M.~Dey, S.~Roy, {Unveiling neutrino mysteries with \ensuremath{\Delta}(27)
  symmetry}, Phys. Lett. B 864 (2025) 139401.
\newblock \href {http://arxiv.org/abs/2309.14769} {\path{arXiv:2309.14769}},
  \href {https://doi.org/10.1016/j.physletb.2025.139401}
  {\path{doi:10.1016/j.physletb.2025.139401}}.

\bibitem{Dinh:2012bp}
D.~N. Dinh, A.~Ibarra, E.~Molinaro, S.~T. Petcov, {The $\mu - e$ Conversion in
  Nuclei, $\mu \to e \gamma, \mu \to 3e$ Decays and TeV Scale See-Saw Scenarios
  of Neutrino Mass Generation}, JHEP 08 (2012) 125, [Erratum: JHEP 09, 023
  (2013)].
\newblock \href {http://arxiv.org/abs/1205.4671} {\path{arXiv:1205.4671}},
  \href {https://doi.org/10.1007/JHEP08(2012)125}
  {\path{doi:10.1007/JHEP08(2012)125}}.

\bibitem{Barrie:2022ake}
N.~D. Barrie, S.~T. Petcov, {Lepton Flavour Violation tests of Type II Seesaw
  Leptogenesis}, JHEP 01 (2023) 001.
\newblock \href {http://arxiv.org/abs/2210.02110} {\path{arXiv:2210.02110}},
  \href {https://doi.org/10.1007/JHEP01(2023)001}
  {\path{doi:10.1007/JHEP01(2023)001}}.

\bibitem{Apollinari:2015bam}
{High-Luminosity Large Hadron Collider (HL-LHC) : Preliminary Design Report}
  (12 2015).
\newblock \href {https://doi.org/10.5170/CERN-2015-005}
  {\path{doi:10.5170/CERN-2015-005}}.

\bibitem{Benedikt:2018ofy}
M.~Benedikt, F.~Zimmermann, {Proton Colliders at the Energy Frontier}, Nucl.
  Instrum. Meth. A 907 (2018) 200--208.
\newblock \href {http://arxiv.org/abs/1803.09723} {\path{arXiv:1803.09723}},
  \href {https://doi.org/10.1016/j.nima.2018.03.021}
  {\path{doi:10.1016/j.nima.2018.03.021}}.

\bibitem{FCC:2018vvp}
A.~Abada, et~al., {FCC-hh: The Hadron Collider}: {Future Circular Collider
  Conceptual Design Report Volume 3}, Eur. Phys. J. ST 228~(4) (2019)
  755--1107.
\newblock \href {https://doi.org/10.1140/epjst/e2019-900087-0}
  {\path{doi:10.1140/epjst/e2019-900087-0}}.

\bibitem{Antusch:2018svb}
S.~Antusch, O.~Fischer, A.~Hammad, C.~Scherb, {Low scale type II seesaw:
  Present constraints and prospects for displaced vertex searches}, JHEP 02
  (2019) 157.
\newblock \href {http://arxiv.org/abs/1811.03476} {\path{arXiv:1811.03476}},
  \href {https://doi.org/10.1007/JHEP02(2019)157}
  {\path{doi:10.1007/JHEP02(2019)157}}.

\bibitem{T2K:2011qtm}
K.~Abe, et~al., {The T2K Experiment}, Nucl. Instrum. Meth. A 659 (2011)
  106--135.
\newblock \href {http://arxiv.org/abs/1106.1238} {\path{arXiv:1106.1238}},
  \href {https://doi.org/10.1016/j.nima.2011.06.067}
  {\path{doi:10.1016/j.nima.2011.06.067}}.

\bibitem{T2K:2025yoy}
K.~Abe, et~al., {Results from the T2K experiment on neutrino mixing including a
  new far detector $\mu$-like sample} (6 2025).
\newblock \href {http://arxiv.org/abs/2506.05889} {\path{arXiv:2506.05889}}.

\bibitem{NOvA:2007rmc}
D.~S. Ayres, et~al., {The NOvA Technical Design Report} (10 2007).
\newblock \href {https://doi.org/10.2172/935497} {\path{doi:10.2172/935497}}.

\bibitem{Catano-Mur:2022kyq}
E.~Catano-Mur, {Recent results from NOvA}, in: {56th Rencontres de Moriond on
  Electroweak Interactions and Unified Theories}, 2022.
\newblock \href {http://arxiv.org/abs/2206.03542} {\path{arXiv:2206.03542}}.

\bibitem{DUNE:2021tad}
V.~Hewes, et~al., {Deep Underground Neutrino Experiment (DUNE) Near Detector
  Conceptual Design Report}, Instruments 5~(4) (2021) 31.
\newblock \href {http://arxiv.org/abs/2103.13910} {\path{arXiv:2103.13910}},
  \href {https://doi.org/10.3390/instruments5040031}
  {\path{doi:10.3390/instruments5040031}}.

\end{thebibliography}
\end{document}